\documentclass[aps,prd,amsmath,floats,floatfix, twocolumn, superscriptaddress,nofootinbib,showpacs,longbibliography]{revtex4-1}
\usepackage[T1]{fontenc}
\usepackage[utf8]{inputenc}
\usepackage{lmodern}
\usepackage{color}
\usepackage{calc}
\usepackage[normalem]{ulem}
\usepackage{amsmath,amssymb,graphicx}
\usepackage{bm}
\usepackage{multirow}
\usepackage{longtable}
\usepackage{microtype}
\usepackage{booktabs}
\usepackage{times}
\usepackage[colorlinks, pdfborder={0 0 0}]{hyperref}
\usepackage{soul}
\definecolor{LinkColor}{rgb}{0.75, 0, 0}
\definecolor{CiteColor}{rgb}{0, 0.5, 0.5}
\definecolor{UrlColor}{rgb}{0, 0, 0.75}
\hypersetup{linkcolor=LinkColor}
\hypersetup{citecolor=CiteColor}
\hypersetup{urlcolor=UrlColor}

\newcommand{\x}{\mathbf{x}}
\newcommand{\y}{\mathbf{y}}
\newcommand{\ba}{\mathbf{a}}
\newcommand{\bb}{\mathbf{b}}

\newcommand{\xit}{\xi_\mathrm{tr}}

\newcommand{\xisdm}{\xi_\mathrm{sm}^\mathrm{m}}

\newcommand{\bmu}{\bm \mu}

\newcommand{\bnu}{\bm \nu}
\newcommand{\Pt}{P_\mathrm{tr}}

\newcommand{\C}{\mathsf{C}}

\newcommand{\ddelta}{\delta^{(3)}}

\begin{document}

\title{Probing the large scale structure using gravitational-wave observations of binary black holes}

\author{Aditya~Vijaykumar}
\affiliation {International Centre for Theoretical Sciences, Tata Institute of Fundamental Research, Bangalore 560089, India }
\author{M.~V.~S.~Saketh}
\affiliation {International Centre for Theoretical Sciences, Tata Institute of Fundamental Research, Bangalore 560089, India }
\affiliation {Indian Institute of Technology, Kanpur, India}
\affiliation {Department of Physics, University of Maryland, College Park, Maryland 20742, USA}
\author{Sumit~Kumar}
\affiliation {Max-Planck-Institut f$\ddot{u}$r Gravitationsphysik (Albert-Einstein-Institut), D-30167 Hannover, Germany }
\affiliation {Leibniz Universit$\ddot{a}$t Hannover, D-30167 Hannover, Germany}
\affiliation {International Centre for Theoretical Sciences, Tata Institute of Fundamental Research, Bangalore 560089, India }
\author{Parameswaran Ajith}
\affiliation {International Centre for Theoretical Sciences, Tata Institute of Fundamental Research, Bangalore 560089, India }
\affiliation {Canadian Institute for Advanced Research,CIFAR Azrieli Global Scholar, MaRS Centre, West Tower, 661 University Ave, Toronto, ON M5G 1M1, Canada}
\author{Tirthankar Roy Choudhury}
\affiliation {National Centre for Radio Astrophysics, Tata Institute of Fundamental Research, Pune, India}

\begin{abstract}
Third generation gravitational-wave (GW) detectors are expected to detect a large number of binary black holes (BBHs) to large redshifts, opening up an independent probe of the large scale structure using their clustering. This probe will be complementary to the probes using galaxy clustering --- GW events could be observed up to very large redshifts ($z \sim 10$) although the source localization will be much poorer at large distances ($\sim$ tens of square degrees). We explore the possibility of probing the large scale structure from the spatial distribution of the observed BBH population, using their two-point (auto)correlation function. We find that we can estimate the bias factor of population of BBH (up to $z \sim 0.7$) with a few years of observations with these detectors. Our method relies solely on the source-location posteriors obtained from the GW events and does not require any information from electromagnetic observations. This will help in identifying the type of galaxies that host the BBH population, thus shedding light on their origins. 
\end{abstract}

\keywords{gravitational waves --- population synthesis --- binary black holes}
\maketitle

\section{Introduction}
Gravitational-wave (GW) observations by LIGO and Virgo have opened a new era of astronomy~\cite{Abbott:2016blz}. On the completion of the third observing run, the LIGO-Virgo-KAGRA collaboration published a combined catalog of GW transients (GWTC-3~\cite{LIGOScientific:2018mvr, LIGOScientific:2020ibl, LIGOScientific:2021djp}), which reported $\sim 90$ significant detections of GWs from compact binaries mergers. Independent analyses of the public LIGO-Virgo data have revealed a few additional events in the same data set~\cite{Venumadhav:2019lyq, Olsen:2022pin,Nitz:2019hdf, Nitz:2021zwj}. Detection of GW events has become a routine now and GW sky is filling up rapidly.

The dominant sources of GW signals for the LIGO-Virgo-KAGRA detectors~\cite{LIGODet,VirgoDet,KAGRADet} is the merger of compact objects. Several hundreds to thousands of such observations are expected in the next few years \cite{Aasi:2013wya}. LIGO-India~\cite{LI-Det, Saleem:2021iwi} is expected to come online sometime during next decade, coinciding with the upgraded Advanced LIGO (A+) detectors~\cite{A+}. These additional detectors will significantly improve the localization of binary mergers. There are several ongoing efforts to build the next generation of ground based detectors. {Proposals for next generation detectors include that of i) LIGO Voyager, which is expected to observe binary neutron stars (BNSs) up to a horizon redshift of $z \sim 0.5$ \cite{LIGO:ISWP2019,Adhikari_2019}, ii) Einstein Telescope (ET), which is expected to have a BNS horizon of $z \sim 2$ \cite{LIGO:ISWP2019,Punturo:2010zz}, and iii) Cosmic Explorer (CE) with an expected BNS horizon of $z \sim 20$ \cite{LIGO:ISWP2019,Dwyer_2015}}. Third generation (3G) detectors like CE and ET will have sensitivity that will be an order of magnitude better than that of Advanced LIGO and will be sensitive to frequencies as low as 1Hz.

Although the network configuration and the sensitivity of the proposed 3G detectors are not finalized, studies of various configurations and their implication on source localization and parameter estimation suggest that these detectors will be able to observe binary black hole (BBH) mergers up to {large redshifts (detection horizon up to $z \sim 100$)~\cite{LIGO:ISWP2019}}. For the redshift range $z \in [0,3] $, a significant fraction of BBH mergers can be localized to within 1 square degree \cite{Vitale:2016icu}. 
These observations will, in essence, create a ``survey'' of well-localized GW events across the sky, akin to galaxy surveys.
Since this survey would trace the underlying large-scale structure of the universe, it is natural to ask what properties of the large-scale structure GW events track, and what that tells us about the underlying astrophysics describing the binaries. In this work, we investigate the possibility of detecting one feature of the large-scale structure (LSS) tracked by GW events, the large-scale bias, using 3G detector network solely from GW observations.

This paper is organized as follows: In section \ref{section:gw_cosmology}, we discuss the current status of cosmology using GW observations. section \ref{section:methodology} describes the methods of probing LSS features using galaxy observations and how can we extend it to GW observations with 3G detectors. In section \ref{section:simulations}, we describe the simulations performed in this study and their results. Section \ref{section:summary} summarizes the results.

\section{Cosmology using GWs}
\label{section:gw_cosmology}
During the second observing run, LIGO and Virgo detected GWs from a BNS merger, GW170817 \cite{GBM:2017lvd}, for the first time. Electromagnetic (EM) counterparts of this event were also detected by several telescopes \cite{Goldstein:2017mmi, LIGOScientific:2017zic, Savchenko:2017ffs}, which enabled the identification of the host galaxy of the merger. This led to a precise measurement of the redshift of GW170817 and the first measurement of the Hubble constant $H_0$ from GW observations \cite{Abbott:2017xzu}. BNS detections expected in the near future with EM counterparts should improve the precision of this measurement, potentially contributing to resolving the apparent tension between the Planck measurement of $H_0$ \cite{Aghanim:2018eyx} and that from type Ia supernovae \cite{Riess:2016jrr}. Some studies also explore the techniques of cross-matching or cross-correlating galaxy catalogs with BBH observations to constrain $H_0$ \cite{Schutz:1986gp,DelPozzo:2011yh,Chen:2017rfc,Fishbach:2018gjp,Nair:2018ign,Osato:2018mtm,Soares-Santos:2019irc,Gray:2019ksv,Abbott:2019yzh}.

With a large number of GW detections expected in the near future, we will have a population of BBH and BNS mergers distributed over a large redshift range, providing a new tracer of the large scale structure. Recent studies show that by cross correlating the GW events with galaxy catalogs, the large scale structure can be probed by estimating the linear bias \cite{Namikawa:2016edr,Calore:2020bpd} or by the lensing of GWs \cite{Mukherjee:2019wcg}. In this work, we explore the possibility of probing the clustering of BBHs by estimating their two-point (auto)correlation function. If these mergers happen in specific types of galaxies, the clustering of the BBHs should trace that of such galaxies. If, for some reason, BBHs are predominantly distributed outside galaxies, their clustering information should reveal this. Thus, an independent estimation of the clustering of BBHs offers an interesting probe of not only the large scale structure, but also the astrophysical environment of the mergers.

\section{Large Scale Structures of the Universe}
\label{section:methodology}
The two-point correlation function (2PCF) $\xi({r})$ is related to the excess probability $\delta P({r})$, above what is expected for a random distribution, of finding a pair of objects (e.g., galaxies or, in the context of this work, BBH mergers) separated by distance ${r}$. This can be expressed as 
\begin{equation}
\delta P({r}) = n \, [1+\xi({r})] \, dV,
\end{equation}
where $n$ is the number of objects per unit volume and $dV$ is the volume element. For the matter overdensity field $\delta(\x) := \rho(\x) / \overline{\rho} - 1$, where $\rho(\x)$ is the local matter density and $\overline{\rho}$ the mean matter density of the Universe, the 2PCF is given by
\begin{equation}
\xi({r}) = \left<\delta(\x) \, \delta(\y)\right> \label{corrfunc},
\end{equation}
where angle brackets denote the ensemble average which, in turn, can be estimated by averaging over a large volume. The above equation assumes statistical homogeneity and isotropy of the universe, hence $\xi$ is only a function of the magnitude $r$ of the separation vector $\y-\x$ between the two points $\x$ and $\y$. In general, the 2PCF is also a function of the redshift $z$. However, when we restrict ourselves to a relatively narrow redshift bin $\Delta z$, it can be assumed to be a constant within that redshift range. 

The distribution of the galaxies in the Universe is expected to trace the underlying matter distribution. At large scales, to a good approximation, the 2PCF of the galaxies $\xi_\mathrm{gal}(r)$ is related to that of matter $\xi_\mathrm{m}(r)$ through a simple relation~\cite{Kaiser:1984sw}
\begin{equation}
\xi_\mathrm{gal}(r) = b_\mathrm{gal}^2 \, \xi_\mathrm{m}(r),
\end{equation}
where $b_\mathrm{gal}$ is the galaxy bias, taken to be scale-independent. Usually, the value of $b_\mathrm{gal}$ depends on the luminosity and color type of galaxies \cite{Zehavi:2004ii}. Similarly, we can also define a bias which quantifies the clustering of the observed BBH population: 
\begin{equation}
\xi_\mathrm{BBH}(r) = b_\mathrm{BBH}^2 \, \xi_\mathrm{m}(r). 
\end{equation}
If we are able to measure $b_\mathrm{BBH}$ from GW observations, this would allow us to compare it against $b_\mathrm{gal}$ estimated from other observations (e.g., EM galaxy surveys), thus providing hints to the host environments of the BBH mergers. 

\subsection{Estimating the BBH correlation function from GW observations}
The interpretation of $ \xi(r) $ as the excess probability of finding points separated by a distance $ r $ allows one to construct fast estimators of the correlation function from data. The Landy-Szalay (LS) estimator \cite{Landy:1993yu} is the most commonly-used estimator, and is given by, 
\begin{equation}
\label{LS-estimator}
\xi(r) = \left[{DD(r)-2DR(r) +RR(r)}\right]~{RR(r)}^{-1} .
\end{equation}
\noindent

\begin{figure}
\includegraphics[width=0.9\columnwidth]{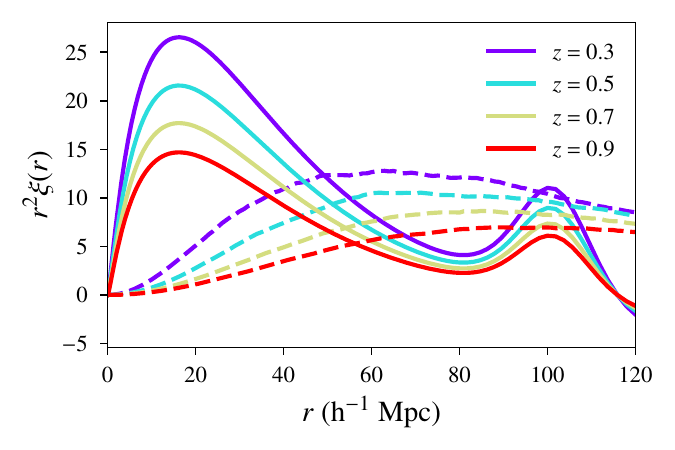}
\caption{The ``smeared'' correlation function (dashed lines) and the ``true'' correlation function (solid lines) for various redshifts. The true correlation function is simply the  matter correlation function calculated using Eisenstein-Hu prescription \cite{Eisenstein:1997jh} for the standard model of cosmology. The ``smearing'' of the correlation function due to measurement errors is calculated assuming that the distribution of errors in localization of GW population follow a Gaussian distribution with mean $\{\mu_\mathrm{RA} = 0.5^\circ, \mu_\mathrm{dec} = 0.5^\circ, \mu_\mathrm{d} = 50~h^{-1}~\mathrm{Mpc}\}$ and standard deviation $\{\sigma_\mathrm{RA} = 0.5^\circ, \sigma_\mathrm{dec} = 0.5^\circ, \sigma_\mathrm{d} = 20 h^{-1}~\mathrm{Mpc}\}$.}
\label{fig:proc_corr_fun}
\end{figure}

\begin{figure}[t]
\centering
\includegraphics[width=0.9\columnwidth]{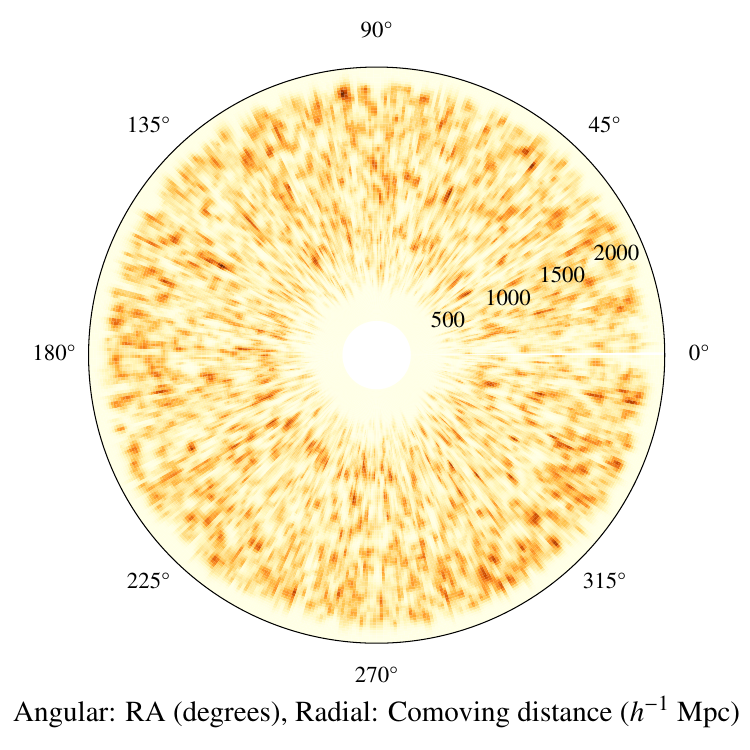}
\caption{An example of probability field obtained from localization posteriors from a realization of simulated catalog of BBH observations in redshift range $z \in [0.1,1.1]$. The radial direction corresponds to comoving distance and the angular direction corresponds to RA (the dec coordinate is projected out). BBH events are distributed according to the input power spectrum and bias factor ($=1.5$) in each redshift bin. The errors on localization are drawn from a probability distribution described in the text.}
\label{fig:bbh_catalog}
\end{figure}

Here, $ DD(r) $ denotes the number of point-pairs in the data (galaxy catalogs or BBH mergers) separated by a distance $r$, $ RR(r) $ denotes the number of point-pairs in an equal-sized simulated random catalog separated by a distance $ r $, and $ DR(r) $ denotes the number of data-random point pairs separated by a distance $ r $. Since the size of the simulated random data set is something we have control over, we can choose to have more number of points in the random catalog. If $ N_D, N_R $ are the number of points in the data and random catalogs respectively, for general $N_D, N_R$,  Eq.\eqref{LS-estimator} gets modified \cite{Landy:1993yu} to,
\begin{equation}
\label{LS-estimator-different-N}
\xi(r) = \left[{\dfrac{DD(r)}{N_D(N_D-1)}-\dfrac{DR(r)}{N_D N_R} +\dfrac{RR(r)}{N_R(N_R-1)}}\right] \, \left[{\dfrac{RR(r)}{N_R(N_R-1)}}\right]^{-1} .
\end{equation}
\begin{figure}
\includegraphics[width=0.9\columnwidth]{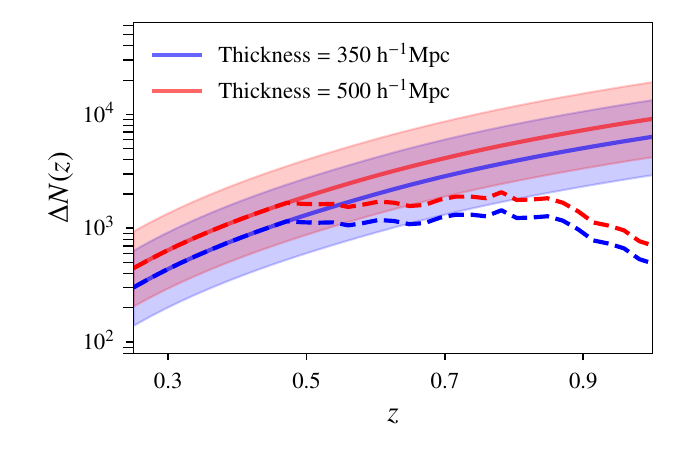}
\caption{Solid curve with shaded region shows the total number of merger events as a function of redshift in shell of thickness $\sim 350~h^{-1}$~Mpc and $\sim 500~h^{-1}$~Mpc in comoving distance. These numbers are calculated by assuming the redshift distribution of BBHs from \cite{Belczynski} and the local merger rates of BBHs estimated in \cite{LIGO-Rates}. The dashed lines show the average number of mergers in the shell of given thickness for which the errors in sky localization are within a degree square and errors in estimating the comoving distance are $\le 90~h^{-1}$~Mpc for a network of three 3G detectors.} 
\label{fig:mergers_events}
\end{figure}
\begin{figure}[t]
\includegraphics[width=0.9\columnwidth]{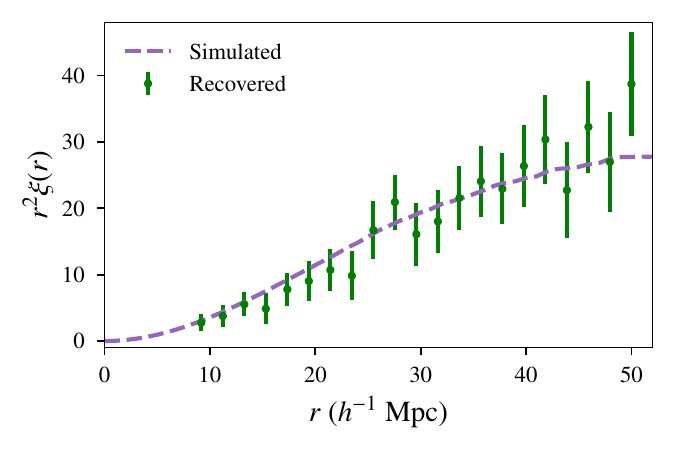}
\caption{Smeared correlation function for a given distribution of localization errors is plotted along with the one recovered from simulated events at redshift 0.3 and input bias factor of 1.5. Smeared correlation function is scaled with input bias for comparison. We used 5000 simulated events distributed in a shell of thickness 350 $h^{-1}$~Mpc around the given redshift. }
\label{fig:simulation_example}
\end{figure}
The correlation function of the galaxies from a survey can be estimated using Eq.\eqref{LS-estimator-different-N}. With next generation GW detectors like ET and CE, we expect to detect BBH mergers up to large redshifts. If the number of detections are sufficiently large, we can use their localization information to study how these GW events are clustered by estimating the correlation function $\xi_\mathrm{BBH}(r)$. 

\subsection{Smearing of the correlation function due to GW localization errors}
The challenge in estimating $\xi_\mathrm{BBH}(r)$ is that the precision in the GW source localization (sky location and distance) will be poor as compared to the galaxy localization (which can effectively be described as a point in the survey volume). Due to the large statistical uncertainties in the GW localization, the observed correlation function of BBHs will be modified from the actual correlation function --- the poor source localization distributes weights from the points of actual location to a smeared field around those points. The ``smearing'' of the correlation function will depend on the distribution of the GW localization uncertainties from the population. 
The smeared correlation function (Figure~\ref{fig:proc_corr_fun}) can be computed by convolving the actual correlation function with the ensemble averaged localization posteriors obtained from GW data. We describe this below.

In the absence of any measurement errors, the probability distribution $\Pt(\bmu)$ of the location $\bmu$ of BBH mergers is given by 
\begin{equation}
\Pt(\bmu) = N^{-1} \sum_i \ddelta (\bmu - \bmu_i),
\label{eq:P_true}
\end{equation}
where $\ddelta$ is the three dimensional Dirac delta function, $\bmu_i$ denotes the three-dimensional location of BBH $i$, and $N$ is the total number of BBHs in the survey volume $V$ such that $\int_V \Pt(\bmu) \, dV_\mu = 1$ ($dV_\mu$ is the volume element in $\mu$; i.e, in Cartesian coordinates $dV_\mu := d^3\mu)$. Density contrast in this field is given by\footnote{The density contrast $\delta_\mathrm{tr}$ is not to be confused with the Dirac delta function $\ddelta$.}
\begin{equation}
\delta_\mathrm{tr}(\bmu) := \Pt(\bmu)/\bar{\Pt} - 1 =  V \Pt(\bmu) - 1. 
\label{eq:delta_true}
\end{equation}
Above, $\bar{\Pt} = 1/V$ denotes the volume-averaged probability density. The correlation function between two points $\bmu$ and $\bnu$ in the field of density contrast is given by 
\begin{equation}
\xit(\bmu, \bnu) := \left< \delta_\mathrm{tr}(\bmu) \, \delta_\mathrm{tr}(\bnu) \right> = V^2 \left< \Pt(\bmu) \, \Pt(\bnu) \right> - 1,
\label{eq:xi_true}
\end{equation}
where $\left<\right>$ denotes ensemble averages. Using Eq.\eqref{eq:P_true}, we can write 
\begin{equation}
 \left< \Pt(\bmu) \, \Pt(\bnu) \right> = N^{-2} \, \left< \sum_{ij} \ddelta(\bmu - \bmu_i) \, \ddelta(\bnu - \bnu_j) \right>. 
\label{eq:Pcorr_true}
\end{equation}
Now we investigate how the true correlation function $\xit(\bmu, \bnu)$ gets smeared by the presence of measurement uncertainties. Assuming that the localization posteriors follow Gaussian distributions,
\begin{eqnarray}
P_i(\x-\bmu_i, \Delta \bmu_i) & = & \frac{1}{\sqrt{(2\pi)^{3} \, |\C_i|}}  \exp\left[-\frac{1}{2}(\x-\bmu_i-\Delta \bmu_i)^T \right. \nonumber \\ 
& \times & \left. \C_i^{-1}(\x-\bmu_i-\Delta \bmu_i)\right] 
\label{eq:posteriorpfield}
\end{eqnarray}
where $\bmu_i$ is the true location of the $i^{th}$ BBH, $\C_i$ is the covariance matrix for the corresponding localization posterior (assumed to be diagonal), and $\Delta \bmu_i$ is the scatter induced by the detector noise. In the absence of systematic biases $\Delta \bmu_i$ will be distributed according to a Gaussian distribution of mean zero and covariance matrix $\C_i$. We now marginalize $P_i(\x-\bmu_i, \Delta \bmu_i)$ over $\Delta \bmu_i$: 
\begin{eqnarray}
P_i(\x-\bmu_i) & = & \int dV_{\Delta \mu} \, P(\Delta \bmu_i) \, P_i(\x-\bmu_i, \Delta \bmu_i).  
\end{eqnarray}
This averaging can be performed on the posterior (as opposed to the final correlation function) since the noise-induced shifts $\Delta \bmu_i$ are uncorrelated with the BBH locations $\bmu_i$. The resulting posterior $P_i(\x-\bmu_i)$ is a Gaussian distribution with mean $\bmu_i$ and covariance matrix $2\C_i$. Using the property of Dirac delta function, $P_i(\x-\bmu_i)$ can be rewritten as 
\begin{equation}
P_i(\x-\bmu_i) = \int dV_\mu \, P_i(\x-\bmu) \,  \ddelta(\bmu - \bmu_i),
\label{eq:Px_dirac}
\end{equation}
and the probability distribution of the location of a population of BBH mergers is given by
\begin{equation}
P(\x)  = N^{-1} \sum_i P_i(\x-\bmu_i).  
\end{equation}
The correlation function between two points $\x$ and $\y$ of this probability field is given by 
\begin{equation}
 \left< P(\x) P(\y) \right> =  N^{-2} \, \left< \sum_{ij} P_i(\x - \bmu_i) \, P_j(\y - \bnu_j) \right>.
\label{eq:Pcorr}
\end{equation}
{Note that each term in the sum over $i,j$ is equal to the joint posterior probability of the BBH mergers $i$ and $j$ to take the positions $\x$ and $\y$, respectively. In a frequentist interpretation, this is equivalent to the joint probability of drawing two samples of $\x$ and $\y$ from the posteriors of the two events. The relation between this and our simulations should be apparent now. We only want to consider correlations between two different BBH mergers; thus, we restrict the sum to $i \neq j$}. Now, Using Eq.\eqref{eq:Px_dirac}, this can be rewritten as 
\begin{eqnarray}
 \left< P(\x) P(\y) \right> & = & N^{-2} \, \left< \sum_{ij} \int dV_\mu  \, P_i(\x-\bmu) \,  \ddelta(\bmu - \bmu_i)  \nonumber  \right. \\ 
    & \times & \left. \int dV_\nu \,  P_j(\y-\bnu) \, \ddelta(\bnu - \bnu_j) \right>. 
\label{eq:Pcorr}
\end{eqnarray}
Now, we make the following assumptions:
\begin{enumerate}
\item Assuming that the posterior distributions are uncorrelated with the actual location of mergers (uniform sky coverage assumption), we can write: $\left<P_{i}(\x-\bmu)P_{j}(\x-\bnu) \, \ddelta(\bmu-\bmu_{i}) \ddelta(\bnu-\bnu_{j})\right> = \left<P_{i}(\x-\bmu)P_{j}(\x-\bnu)\right> \, \left<\ddelta(\bmu-\bmu_{i})\ddelta(\bnu-\bnu_{j})\right>$

\item Since $P_i$ and $P_j$ are posterior probability distributions estimated from two independent GW events (uncorrelated noise), $\left< P_i(\x-\bmu) P_j(\y-\bnu) \right> = \left< P_i(\x-\bmu) \right> \, \left< P_j(\y-\bnu) \right>$.
\item Motivated by the homogeneity of space, we assume $\left< P_i(\x-\bmu) \right> = P(\x-\bmu)$ and $\left< P_j(\y-\bnu) \right>  = P(\y-\bnu)$. 
\end{enumerate}
Using these assumptions Eq.\eqref{eq:Pcorr} can be rewritten as 
\begin{eqnarray}
 \left< P(\x) P(\y) \right>  =  N^{-2} \, \int dV_\mu \int dV_\nu  \, P(\x-\bmu) \, P(\y-\bnu) \nonumber \\ 
     \times  \left<  \sum_{ij} \ddelta(\bmu - \bmu_i) \, \ddelta(\bnu - \bnu_j) \right>. \nonumber  \\ 
     =  \int dV_\mu \int dV_\nu  \, P(\x-\bmu) \, P(\y-\bnu) \left< \Pt(\bmu) \, \Pt(\bnu) \right>,
\label{eq:Pcorr_v2}
\end{eqnarray}
where we have used Eq.\eqref{eq:Pcorr_true} for the last step. The smeared correlation function of the probability density contrast field $\delta_P(\x) := P(\x)/\bar{P} - 1$ is given by 
\begin{equation}
\xi(\x, \y) = \left< \delta_P(\x) \, \delta_P(\y) \right> = V^2 \left< P(\x) \, P(\y) \right> - 1. 
\label{eq:xi}
\end{equation}
Using Eqs.\eqref{eq:Pcorr_v2} and \eqref{eq:xi_true}, this can be rewritten as 
\begin{equation}
\xi(\x, \y) = \int_V dV_\mu \int_V dV_\nu \,  P(\x-\bmu) \, P(\y-\bnu) \, \xit(\bmu, \bnu)
\label{eq:xi_v2}
\end{equation}
This can be used to compute the smeared correlation function $\xi(\x, \y)$ from the true correlation function $\xit(\bmu, \bnu)$. Essentially we convolve the true correlation function $\xit$ by a smoothing function (ensemble-averaged localization posteriors). 

Due to the homogeneity and isotropy of space, the true correlation function only depends on the magnitude of the difference of its arguments $\xit(|\bnu-\bmu|)$. We  can exploit this by transforming to new integration variables, $\ba:=\x-\bmu$ and $\bb:=\y-\bnu$ to get:
\begin{equation}
\xi(\y-\x) = \int_V dV_a \int_V dV_b \,  P(\ba) \, P(\bb) \, \xit(s) ,
    \label{sfinale}
\end{equation}
where $s := |\bnu-\bmu| = |(\y-\x)-(\bb-\ba)|$. This shows that the smeared correlation function also depends only on the separation of points $\x$ and $\y$, i.e., $\xi(\y-\x)$. However, unlike the true correlation function, the direction also matters unless posterior functions are isotropic. In the case of BBH mergers, we expect the radial uncertainty to be much larger than the angular uncertainties and therefore we cannot demand isotropy and thus the orientation of $\x-\y$ matters. To deal with this, we average Eq.(\ref{sfinale}) over  all orientations for a given $r := |\x-\y|$ to find the spherically averaged correlation function $\xi(r)$. We choose the volume of interest to be large enough to permit every possible orientation with minimal bias.

In this work, we simulate the combined probability distribution of BBHs by placing GW posteriors around the true BBH locations, after introducing a noise-induced scatter in the mean of the posteriors. The posterior distributions of right ascension (RA), declination (dec) and {comoving distance} ($d$), estimated from $N$ simulated events are combined to create a normalized combined posterior probability field $P(\x) = N^{-1} \, \sum_{i=1}^N\, P_i(\x)$, where $\x = \{\mathrm{RA, dec}, d\}$. Assuming that the localization posteriors follow Gaussian distributions,
\begin{equation}
P_i(\x) = \mathcal{N} ~ \exp\left[-\frac{1}{2}(\x-\bmu_i-\Delta \bmu_i)^T\C_i^{-1}(\x-\bmu_i-\Delta \bmu_i)\right] 
\label{eq:posteriorpfield}
\end{equation}
where $\bmu_i$ is the true location of the $i^{th}$ BBH and $\C_i$ is the covariance matrix of the corresponding localization posterior (assumed to be diagonal), while $\mathcal{N} = 1/{\sqrt{(2\pi)^{3} \, |\C_i|}}$ is a normalization constant. Note that the individual posteriors will in general not be centered around the true BBH locations, because of the scatter $\Delta \bmu_i$ introduced by the detector noise. This random scatter is drawn from a mean-zero Gaussian distribution of covariance matrix $\C_i$. Figure~\ref{fig:bbh_catalog} shows the $P(\x)$ from a simulated catalog of BBH observations. 

\section{Simulations and results}
\label{section:simulations}
To put our method to test, we use the publicly available code \textsc{lognormal\_galaxies} \cite{Agrawal:2017khv}, to simulate galaxy catalogs at various redshifts with input power spectrum taken as the matter power spectrum approximated by the fitting function of Eisenstein and Hu \cite{Eisenstein:1997jh} and consistent with the Planck-18 cosmological parameters \cite{Aghanim:2018eyx}. This code enables one to generate mock galaxy catalogs assuming lognormal probability density function for the matter field and galaxies. We assume that GW events occur in any random subsample of the galaxies in the catalog, which essentially implies $b_\mathrm{BBH}=b_\mathrm{gal}$. We simulate three different catalogs having input linear bias $b_\mathrm{gal}=[1.2,1.5,2.0]$. Since it is possible to directly infer the distance (within the localization errors) to the BBH from the GW observations, peculiar velocities of galaxies will not play any role (unlike EM galaxy surveys where the distance is inferred from the redshift). Hence, while generating the catalogs, we switched off peculiar velocities in the code. 

We then simulate the mock BBH catalogs using the steps outlined below and check whether we are able to recover the bias consistent with the input value. For simplicity, we have assumed the input $b_\mathrm{gal}$ to be redshift-independent, however, our conclusions on the recovery of the bias would remain unchanged even if we use an evolving bias. These are the steps involved: 
	
\begin{figure*}[t]
\centering
\includegraphics[scale=.525]{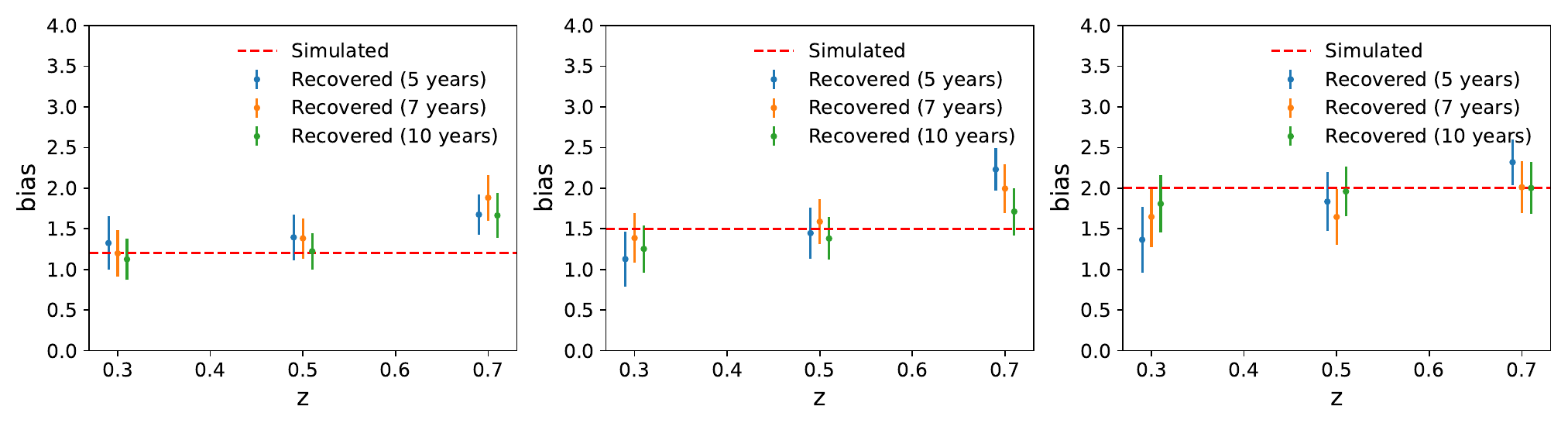} 
\caption{The recovered bias factor $b_\mathrm{BBH}$ from various redshifts bins (with shell thickness of $\sim 350~h^{-1}$ Mpc). The catalogs were created using the matter power spectrum of Eisenstein-Hu with different values of linear bias 1.2 (left) 1.5 (middle) and 2.0 (right). Each subplot shows the estimated bias factor, along with the corresponding error bars ({68\%} confidence regions), using GW observations of BBHs over a period of 5, 7 and 10 years.}
\label{fig:resultsfigure}
\end{figure*}

\begin{enumerate}
\item Choose a shell of thickness $350~h^{-1}$~Mpc around the given redshift. The value was chosen so that we have enough events in the shell, and the actual correlation function does not vary appreciably within the redshift bin. The extent $\Delta z$ of the redshift bin corresponding to this shell thickness at redshift $z = 0.3~ (1.0)$ turns out to be $0.13~ (0.2)$\footnote{An admittedly artificial by-product of this procedure is that we cannot model the redshift evolution of the clustering within a given shell; perhaps a more physically consistent choice would have been to use a lightcone included the evolution of the correlation function. We resort to the former approach since it naturally enables the creation of an all-sky catalog, even though it doesn’t include the full physical content.
}.
\item Randomly select $N$ galaxies from this shell as proxy for GW events and put localization error bars on each event assuming Gaussian posteriors (see below for details regarding uncertainties). 
\item {Select one point from each of the $N$ posteriors. This simulates a particular realization of galaxy locations. Use LS estimator to estimate the correlation function. We repeat this process 1000 times and take the average to get $\xi(r)$.}
\item {To estimate the variance, we create 50 galaxy catalogs corresponding to different realization of the cosmic matter field to account for cosmic variance. For each of these catalogs, we select 20 sub-catalogs $N$ of random galaxies each to account for fluctuations due to sampling, thus amounting to a total of 1000 sub-catalogs. One sub-catalog catalog was taken as realization of our universe and $\xi_\mathrm{BBH}(r)$ was estimated using steps described above. Error bars on $\xi_\mathrm{BBH}(r)$ was placed making use of the scatter estimated from other sub-catalogs.}

\begin{figure}
\includegraphics[width=0.9\columnwidth]{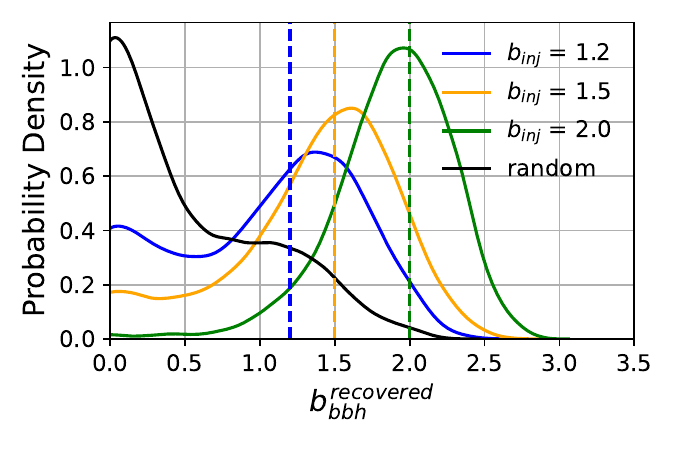}
\caption{Blue, orange, and green curves represent the distribution of stacked posterior from  realizations of simulated universe with injected values of bias are $b_{inj}=1.2,1.5$, and $2.0$ respectively. Black curve represent the same from Universes with random distribution of BBH mergers. This comparison is shown with Universe simulated at the redshift of 0.5 with 10 years of accumulated data.}
\label{fig:random_catalog_comparison}
\end{figure}

\item Estimate the bias factor $b_\mathrm{BBH}$ by comparing the recovered correlation function $\xi_\mathrm{BBH}(r)$ with the smeared matter correlation function $\xisdm(r)$. We estimate the correlation function in the range of comoving distance $r \in [10,50]\sim h^{-1}$ Mpc as this is well within the chosen shell thickness and the linear bias approximation is valid in this range. To find the fit for the bias factor we then define a $\chi^2$ function,
\begin{equation}
\chi^2(b) = \Delta X^T \Sigma^{-1} \Delta X
\end{equation}
where $\Delta X_i = \xi_\mathrm{est}(r_i) - b^2\, \xisdm(r_i)$,  $\xisdm(r_i)$ is the smeared matter correlation function for the given distribution of localization errors, $b$ is the bias factor, $\xi_\mathrm{est}(r_i)$ is the correlation function estimated from the simulated catalog for each $r$-bin $r_i$, and $\Sigma_{ij}$ is the covariance matrix between $i$-th and $j$-th bin. {We estimate the covariance matrix using 1000 sub-catalogs we generated for this study.} We define the likelihood function as,
\begin{equation}
    \mathcal{L}(b) = \exp(-\chi^2(b)/2), \label{eqn:likelihood}
\end{equation}
\noindent
and use Bayesian analysis to estimate posterior distribution corresponding to the likelihood function Eq.~\eqref{eqn:likelihood} for a uniform prior for parameter $b$ in range $b \in [0, 5]$. We use the open source nested sampling based sampler \textsc{dynesty} \cite{2019S&C....29..891H} for this purpose.

\end{enumerate}
Since we used the galaxies as proxy for GW events, we expect the recovered $b_\mathrm{BBH}$ to be consistent with the input bias factor used for simulating the galaxy distribution. Clearly, this method will be valid only when the errors in localization do not exceed the range of comoving distances we are trying to probe. This translates into the requirement that errors in RA, dec should be within a degree and errors in the comoving distances should not exceed few tens of Mpc. To find if this requirement can be fulfilled with 3G detectors, we perform GW parameter estimation studies using a population of BBH events distributed up to redshift $\sim 1.2$ with 3G detector network 2CE-ET (CE locations: one in USA and one in Australia. ET location: proposed one in Europe).  In Table \ref{tab:3g_detectors_location}, we list the location and and low frequency cutoff used for 3G detectors network.
In our simulations, we use BBH population with the powerlaw plus a peak distribution of primary masses $p(m_1) \propto m_{1}^{-\alpha}$ with $\alpha=2.3$ \cite{LIGOScientific:2018jsj}.
We use the \textsc{IMRPhenomPv2} \cite{Hannam:2013oca} waveform available in the \textsc{LALSuite} \cite{lalsuite} software package along with the appropriate detector PSDs \cite{Hild:2010id,Evans:2016mbw} to simulate our signals, and use the \textsc{PyCBCInference} package \cite{Biwer:2018osg} to determine distribution of localization errors. We find that a significant fraction of events up to $z \simeq 1$ fulfills this requirement. Figure \ref{fig:mergers_events} shows number of expected BBH mergers (using the BBH merger rate given in \cite{LIGOScientific:2018jsj}) at various redshifts for one year of observations along with fraction of events that are expected to be localized well enough for this type of study. In our simulations, this selection introduces no significant biases; however possible selection effects need to be considered for the actual analysis. The distribution of the widths of the {68\% credible regions} of the marginalized posteriors on RA, dec and comoving distance can be approximated by truncated Gaussian distributions with mean $\{\mu_\mathrm{RA} = 0.5^\circ, \mu_\mathrm{dec} = 0.5^\circ, \mu_\mathrm{d} = 50~h^{-1}~\mathrm{Mpc}\}$ and standard deviation $\{\sigma_\mathrm{RA} = 0.5^\circ, \sigma_\mathrm{dec} = 0.5^\circ, \sigma_\mathrm{d} = 20 h^{-1}~\mathrm{Mpc}\}$. {For RA and dec, the range of truncated Gaussian was taken to be $\in [0.1^\circ, 1.5^\circ]$ and for comoving distance $\in [20,90]~h^{-1}$~Mpc.} We neglect the correlations between the errors in RA, dec and distance.

\begin{table}
    \caption{The specifications of 3G detector network (location, low frequency cutoff $f_\mathrm{low}$) considered in this study. We use the design sensitivity noise curves for ET and CE. These detectors configuration for CE and ET are also used in previous works \citep{2019CQGra..36v5002H, Nitz2021premerger, Kumar:2021aog}.
    }
    \label{table:detectors}
%\begin{center}
\begin{tabular}{lllcc}
Observatory  & $f_{\textrm{low}}$ &  Latitude & Longitude \\ \hline
Cosmic Explorer USA & 5.2 & 40.8 & -113.8    \\
Cosmic Explorer Australia & 5.2 &-31.5 & 118.0   \\
Einstein Telescope & 2 & 43.6 & 10.5  \\
\label{tab:3g_detectors_location}
\end{tabular}

%\end{center}
\end{table}

Figure~\ref{fig:simulation_example} shows the smeared correlation function compared to the estimated correlation function from a simulation using 5000 GW observation in a shell around the redshift $z=0.3$. Figure~\ref{fig:resultsfigure} shows the bias factor recovered from different redshift bins using different observation duration (5, 7 and 10 years). The estimated $b_\mathrm{BBH}$, in general, are consistent with the simulated bias within error-bars. The small number of events where the actual value is outside the error bars is consistent with statistical fluctuations. Note that even with a moderate observational time of five years, we can recover the bias to within $\sim 20\%$ for $z \lesssim 0.7$. Due to large spread of localization volumes, the bias recovery becomes difficult for high redshift~\footnote{Note that the LS estimator might not be the optimal estimator in the presence of measurement errors like in the case of GW observations. We are investigating alternative methods for the estimation of correlation function in such cases.}. We would like to point out that if the localization volumes with 3G detectors can be improved by either better sensitivity or improvements in waveform modelling, the results presented in this study will improve. 

In order to test the robustness of this method, we simulated ~1000 catalogues with galaxies distributed with $b_{gal}=1.2$, $1.5$, and $2.0$. For reference, we also generated a catalog with no underlying correlation function i.e. galaxies are distributed randomly (corresponding to $b=0$) at redshift $z=0.5$. In Figure \ref{fig:random_catalog_comparison}, we show the stacked posterior samples from all the individual runs. The width of stacked posterior samples depends on the localization volume distribution, and sampling errors. We note that the stacked posteriors are peaked around the injected values. These reference distribution, for given localization volumes, can be used to assign significance to a particular bias recovery measurement with respect to the random distribution.

There are various ways one can use the recovered $b_\mathrm{BBH}(z)$ to understand the properties of the event hosts. For example, one can compare the clustering properties of the galaxies as measured from the optical surveys with $b_\mathrm{BBH}(z)$ and obtain insights on the type of galaxies that host these merger events. Further, the recovered bias can also be related to the host dark matter halo mass \cite{Cooray:2002dia}. In general, if one assumes that the typical masses of the haloes hosting these GW events do not evolve with redshift, one can predict the redshift-dependence of $b_\mathrm{BBH}(z)$ for a given cosmological model. This then can be compared with the observations to understand the formation channels of the BBHs. 

{We would like to emphasize that in order to convert luminosity distance samples obtained from GW localization volumes, to either comoving volume, or redshift, we assume an underlying cosmological model: $\Lambda$CDM model with Planck 2018~\cite{Aghanim:2018eyx} values as implemented in \textsc{AstroPy} \cite{Robitaille:2013mpa,Price-Whelan:2018hus}. The results presented herein are intrinsically tied to this chosen cosmological model. In order to pursue a model independent approach, one needs to marginalize over cosmological parameters. Furthermore, when dealing with real-world data, additional factors and effects come into play. These may include the development of optimal methodologies for estimating the covariance matrix between distance bins, incorporation of detector response functions, consideration of weak lensing effects, and more. We intend to investigate these effects in our future work.}

\section{Summary}
\label{section:summary}
In this work we explored the possibility of probing large scale structure with BBH observations using third generation GW detectors. We showed that bias factor can be estimated using clustering information of BBH events with 5-10 years of observations. This can be achieved solely from the GW observations, without requiring EM counterparts or galaxy catalogs. The bias factor $b_\mathrm{BBH}$ estimated from various redshifts will enable us to find whether the BBH mergers track the distribution of specific types of galaxies, or dark matter halos. Although the statistical precision of the estimated bias $b_{\rm BBH}$ is weaker than that of the galaxy bias obtained from EM galaxy surveys, it is important to note that the GW-based analysis probes the underlying matter distribution using a novel astrophysical tracer, thus enabling an independent probe of the large scale structure. We intend to extend this analysis to include effects such as selection bias, and method of cross correlating with galaxy catalogs to probe higher redshift, etc.

\section{Acknowledgments}
\label{section:acknowledgement}
We thank Jonathan Gair for very useful comments on the manuscript. We also thank Surhud More, Collin Capano, Badri Krishnan, Bala Iyer, Shasvath Kapadia, Archisman Ghosh and David Keitel for useful discussions and comments. Our research is supported by the Department of Atomic Energy, Government of India. SK acknowledges the funding from national post doctoral fellowship (PDF/2016/001294) by Scientific and Engineering Research Board, Govt. of India. In addition, PA's research was supported by the Max Planck Society through a Max Planck Partner Group at ICTS-TIFR and by the Canadian Institute for Advanced Research through the CIFAR Azrieli Global Scholars program. This works makes use of \textsc{NumPy} \cite{Harris:2020xlr}, \textsc{SciPy} \cite{Virtanen:2019joe}, \textsc{Matplotlib} \cite{Hunter:2007}, \textsc{AstroPy} \cite{Robitaille:2013mpa,Price-Whelan:2018hus}, and \textsc{dynesty} \cite{2019S&C....29..891H} software packages. Computations were performed at the Alice cluster at ICTS-TIFR and Atlas cluster at AEI Hannover.

\bibliography{references}

\begin{thebibliography}{62}%
\makeatletter
\providecommand \@ifxundefined [1]{%
 \@ifx{#1\undefined}
}%
\providecommand \@ifnum [1]{%
 \ifnum #1\expandafter \@firstoftwo
 \else \expandafter \@secondoftwo
 \fi
}%
\providecommand \@ifx [1]{%
 \ifx #1\expandafter \@firstoftwo
 \else \expandafter \@secondoftwo
 \fi
}%
\providecommand \natexlab [1]{#1}%
\providecommand \enquote  [1]{``#1''}%
\providecommand \bibnamefont  [1]{#1}%
\providecommand \bibfnamefont [1]{#1}%
\providecommand \citenamefont [1]{#1}%
\providecommand \href@noop [0]{\@secondoftwo}%
\providecommand \href [0]{\begingroup \@sanitize@url \@href}%
\providecommand \@href[1]{\@@startlink{#1}\@@href}%
\providecommand \@@href[1]{\endgroup#1\@@endlink}%
\providecommand \@sanitize@url [0]{\catcode `\\12\catcode `\$12\catcode
  `\&12\catcode `\#12\catcode `\^12\catcode `\_12\catcode `\%12\relax}%
\providecommand \@@startlink[1]{}%
\providecommand \@@endlink[0]{}%
\providecommand \url  [0]{\begingroup\@sanitize@url \@url }%
\providecommand \@url [1]{\endgroup\@href {#1}{\urlprefix }}%
\providecommand \urlprefix  [0]{URL }%
\providecommand \Eprint [0]{\href }%
\providecommand \doibase [0]{http://dx.doi.org/}%
\providecommand \selectlanguage [0]{\@gobble}%
\providecommand \bibinfo  [0]{\@secondoftwo}%
\providecommand \bibfield  [0]{\@secondoftwo}%
\providecommand \translation [1]{[#1]}%
\providecommand \BibitemOpen [0]{}%
\providecommand \bibitemStop [0]{}%
\providecommand \bibitemNoStop [0]{.\EOS\space}%
\providecommand \EOS [0]{\spacefactor3000\relax}%
\providecommand \BibitemShut  [1]{\csname bibitem#1\endcsname}%
\let\auto@bib@innerbib\@empty
%</preamble>
\bibitem [{\citenamefont {Abbott}\ \emph {et~al.}(2016)\citenamefont {Abbott}
  \emph {et~al.}}]{Abbott:2016blz}%
  \BibitemOpen
  \bibfield  {author} {\bibinfo {author} {\bibfnamefont {B.~P.}\ \bibnamefont
  {Abbott}} \emph {et~al.} (\bibinfo {collaboration} {LIGO Scientific,
  Virgo}),\ }\bibfield  {title} {\enquote {\bibinfo {title} {{Observation of
  Gravitational Waves from a Binary Black Hole Merger}},}\ }\href {\doibase
  10.1103/PhysRevLett.116.061102} {\bibfield  {journal} {\bibinfo  {journal}
  {Phys. Rev. Lett.}\ }\textbf {\bibinfo {volume} {116}},\ \bibinfo {pages}
  {061102} (\bibinfo {year} {2016})},\ \Eprint
  {http://arxiv.org/abs/1602.03837} {arXiv:1602.03837 [gr-qc]} \BibitemShut
  {NoStop}%
%%CITATION = ARXIV:1602.03837;%%
\bibitem [{\citenamefont {Abbott}\ \emph
  {et~al.}(2019{\natexlab{a}})\citenamefont {Abbott} \emph
  {et~al.}}]{LIGOScientific:2018mvr}%
  \BibitemOpen
  \bibfield  {author} {\bibinfo {author} {\bibfnamefont {B.~P.}\ \bibnamefont
  {Abbott}} \emph {et~al.} (\bibinfo {collaboration} {LIGO Scientific,
  Virgo}),\ }\bibfield  {title} {\enquote {\bibinfo {title} {{GWTC-1: A
  Gravitational-Wave Transient Catalog of Compact Binary Mergers Observed by
  LIGO and Virgo during the First and Second Observing Runs}},}\ }\href
  {\doibase 10.1103/PhysRevX.9.031040} {\bibfield  {journal} {\bibinfo
  {journal} {Phys. Rev. X}\ }\textbf {\bibinfo {volume} {9}},\ \bibinfo {pages}
  {031040} (\bibinfo {year} {2019}{\natexlab{a}})},\ \Eprint
  {http://arxiv.org/abs/1811.12907} {arXiv:1811.12907 [astro-ph.HE]}
  \BibitemShut {NoStop}%
\bibitem [{\citenamefont {Abbott}\ \emph
  {et~al.}(2021{\natexlab{a}})\citenamefont {Abbott} \emph
  {et~al.}}]{LIGOScientific:2020ibl}%
  \BibitemOpen
  \bibfield  {author} {\bibinfo {author} {\bibfnamefont {R.}~\bibnamefont
  {Abbott}} \emph {et~al.} (\bibinfo {collaboration} {LIGO Scientific,
  Virgo}),\ }\bibfield  {title} {\enquote {\bibinfo {title} {{GWTC-2: Compact
  Binary Coalescences Observed by LIGO and Virgo During the First Half of the
  Third Observing Run}},}\ }\href {\doibase 10.1103/PhysRevX.11.021053}
  {\bibfield  {journal} {\bibinfo  {journal} {Phys. Rev. X}\ }\textbf {\bibinfo
  {volume} {11}},\ \bibinfo {pages} {021053} (\bibinfo {year}
  {2021}{\natexlab{a}})},\ \Eprint {http://arxiv.org/abs/2010.14527}
  {arXiv:2010.14527 [gr-qc]} \BibitemShut {NoStop}%
\bibitem [{\citenamefont {Abbott}\ \emph
  {et~al.}(2021{\natexlab{b}})\citenamefont {Abbott} \emph
  {et~al.}}]{LIGOScientific:2021djp}%
  \BibitemOpen
  \bibfield  {author} {\bibinfo {author} {\bibfnamefont {R.}~\bibnamefont
  {Abbott}} \emph {et~al.} (\bibinfo {collaboration} {LIGO Scientific, VIRGO,
  KAGRA}),\ }\bibfield  {title} {\enquote {\bibinfo {title} {{GWTC-3: Compact
  Binary Coalescences Observed by LIGO and Virgo During the Second Part of the
  Third Observing Run}},}\ }\href@noop {} {\  (\bibinfo {year}
  {2021}{\natexlab{b}})},\ \Eprint {http://arxiv.org/abs/2111.03606}
  {arXiv:2111.03606 [gr-qc]} \BibitemShut {NoStop}%
\bibitem [{\citenamefont {Venumadhav}\ \emph {et~al.}(2020)\citenamefont
  {Venumadhav}, \citenamefont {Zackay}, \citenamefont {Roulet}, \citenamefont
  {Dai},\ and\ \citenamefont {Zaldarriaga}}]{Venumadhav:2019lyq}%
  \BibitemOpen
  \bibfield  {author} {\bibinfo {author} {\bibfnamefont {Tejaswi}\ \bibnamefont
  {Venumadhav}}, \bibinfo {author} {\bibfnamefont {Barak}\ \bibnamefont
  {Zackay}}, \bibinfo {author} {\bibfnamefont {Javier}\ \bibnamefont {Roulet}},
  \bibinfo {author} {\bibfnamefont {Liang}\ \bibnamefont {Dai}}, \ and\
  \bibinfo {author} {\bibfnamefont {Matias}\ \bibnamefont {Zaldarriaga}},\
  }\bibfield  {title} {\enquote {\bibinfo {title} {{New binary black hole
  mergers in the second observing run of Advanced LIGO and Advanced Virgo}},}\
  }\href {\doibase 10.1103/PhysRevD.101.083030} {\bibfield  {journal} {\bibinfo
   {journal} {Phys. Rev. D}\ }\textbf {\bibinfo {volume} {101}},\ \bibinfo
  {pages} {083030} (\bibinfo {year} {2020})},\ \Eprint
  {http://arxiv.org/abs/1904.07214} {arXiv:1904.07214 [astro-ph.HE]}
  \BibitemShut {NoStop}%
\bibitem [{\citenamefont {Olsen}\ \emph {et~al.}(2022)\citenamefont {Olsen},
  \citenamefont {Venumadhav}, \citenamefont {Mushkin}, \citenamefont {Roulet},
  \citenamefont {Zackay},\ and\ \citenamefont {Zaldarriaga}}]{Olsen:2022pin}%
  \BibitemOpen
  \bibfield  {author} {\bibinfo {author} {\bibfnamefont {Seth}\ \bibnamefont
  {Olsen}}, \bibinfo {author} {\bibfnamefont {Tejaswi}\ \bibnamefont
  {Venumadhav}}, \bibinfo {author} {\bibfnamefont {Jonathan}\ \bibnamefont
  {Mushkin}}, \bibinfo {author} {\bibfnamefont {Javier}\ \bibnamefont
  {Roulet}}, \bibinfo {author} {\bibfnamefont {Barak}\ \bibnamefont {Zackay}},
  \ and\ \bibinfo {author} {\bibfnamefont {Matias}\ \bibnamefont
  {Zaldarriaga}},\ }\bibfield  {title} {\enquote {\bibinfo {title} {{New binary
  black hole mergers in the LIGO-Virgo O3a data}},}\ }\href {\doibase
  10.1103/PhysRevD.106.043009} {\bibfield  {journal} {\bibinfo  {journal}
  {Phys. Rev. D}\ }\textbf {\bibinfo {volume} {106}},\ \bibinfo {pages}
  {043009} (\bibinfo {year} {2022})},\ \Eprint
  {http://arxiv.org/abs/2201.02252} {arXiv:2201.02252 [astro-ph.HE]}
  \BibitemShut {NoStop}%
\bibitem [{\citenamefont {Nitz}\ \emph {et~al.}(2020)\citenamefont {Nitz},
  \citenamefont {Dent}, \citenamefont {Davies}, \citenamefont {Kumar},
  \citenamefont {Capano}, \citenamefont {Harry}, \citenamefont {Mozzon},
  \citenamefont {Nuttall}, \citenamefont {Lundgren},\ and\ \citenamefont
  {T\'apai}}]{Nitz:2019hdf}%
  \BibitemOpen
  \bibfield  {author} {\bibinfo {author} {\bibfnamefont {Alexander~H.}\
  \bibnamefont {Nitz}}, \bibinfo {author} {\bibfnamefont {Thomas}\ \bibnamefont
  {Dent}}, \bibinfo {author} {\bibfnamefont {Gareth~S.}\ \bibnamefont
  {Davies}}, \bibinfo {author} {\bibfnamefont {Sumit}\ \bibnamefont {Kumar}},
  \bibinfo {author} {\bibfnamefont {Collin~D.}\ \bibnamefont {Capano}},
  \bibinfo {author} {\bibfnamefont {Ian}\ \bibnamefont {Harry}}, \bibinfo
  {author} {\bibfnamefont {Simone}\ \bibnamefont {Mozzon}}, \bibinfo {author}
  {\bibfnamefont {Laura}\ \bibnamefont {Nuttall}}, \bibinfo {author}
  {\bibfnamefont {Andrew}\ \bibnamefont {Lundgren}}, \ and\ \bibinfo {author}
  {\bibfnamefont {M\'arton}\ \bibnamefont {T\'apai}},\ }\bibfield  {title}
  {\enquote {\bibinfo {title} {{2-OGC: Open Gravitational-wave Catalog of
  binary mergers from analysis of public Advanced LIGO and Virgo data}},}\
  }\href {\doibase 10.3847/1538-4357/ab733f} {\bibfield  {journal} {\bibinfo
  {journal} {Astrophys. J.}\ }\textbf {\bibinfo {volume} {891}},\ \bibinfo
  {pages} {123} (\bibinfo {year} {2020})},\ \Eprint
  {http://arxiv.org/abs/1910.05331} {arXiv:1910.05331 [astro-ph.HE]}
  \BibitemShut {NoStop}%
\bibitem [{\citenamefont {Nitz}\ \emph {et~al.}(2023)\citenamefont {Nitz},
  \citenamefont {Kumar}, \citenamefont {Wang}, \citenamefont {Kastha},
  \citenamefont {Wu}, \citenamefont {Sch\"afer}, \citenamefont {Dhurkunde},\
  and\ \citenamefont {Capano}}]{Nitz:2021zwj}%
  \BibitemOpen
  \bibfield  {author} {\bibinfo {author} {\bibfnamefont {Alexander~H.}\
  \bibnamefont {Nitz}}, \bibinfo {author} {\bibfnamefont {Sumit}\ \bibnamefont
  {Kumar}}, \bibinfo {author} {\bibfnamefont {Yi-Fan}\ \bibnamefont {Wang}},
  \bibinfo {author} {\bibfnamefont {Shilpa}\ \bibnamefont {Kastha}}, \bibinfo
  {author} {\bibfnamefont {Shichao}\ \bibnamefont {Wu}}, \bibinfo {author}
  {\bibfnamefont {Marlin}\ \bibnamefont {Sch\"afer}}, \bibinfo {author}
  {\bibfnamefont {Rahul}\ \bibnamefont {Dhurkunde}}, \ and\ \bibinfo {author}
  {\bibfnamefont {Collin~D.}\ \bibnamefont {Capano}},\ }\bibfield  {title}
  {\enquote {\bibinfo {title} {{4-OGC: Catalog of Gravitational Waves from
  Compact Binary Mergers}},}\ }\href {\doibase 10.3847/1538-4357/aca591}
  {\bibfield  {journal} {\bibinfo  {journal} {Astrophys. J.}\ }\textbf
  {\bibinfo {volume} {946}},\ \bibinfo {pages} {59} (\bibinfo {year} {2023})},\
  \Eprint {http://arxiv.org/abs/2112.06878} {arXiv:2112.06878 [astro-ph.HE]}
  \BibitemShut {NoStop}%
\bibitem [{\citenamefont {Aasi}\ \emph {et~al.}(2015)\citenamefont {Aasi} \emph
  {et~al.}}]{LIGODet}%
  \BibitemOpen
  \bibfield  {author} {\bibinfo {author} {\bibfnamefont {J.}~\bibnamefont
  {Aasi}} \emph {et~al.} (\bibinfo {collaboration} {LIGO Scientific}),\
  }\bibfield  {title} {\enquote {\bibinfo {title} {{Advanced LIGO}},}\ }\href
  {\doibase 10.1088/0264-9381/32/7/074001} {\bibfield  {journal} {\bibinfo
  {journal} {Class. Quant. Grav.}\ }\textbf {\bibinfo {volume} {32}},\ \bibinfo
  {pages} {074001} (\bibinfo {year} {2015})},\ \Eprint
  {http://arxiv.org/abs/1411.4547} {arXiv:1411.4547 [gr-qc]} \BibitemShut
  {NoStop}%
%%CITATION = ARXIV:1411.4547;%%
\bibitem [{\citenamefont {Acernese}\ \emph {et~al.}(2015)\citenamefont
  {Acernese} \emph {et~al.}}]{VirgoDet}%
  \BibitemOpen
  \bibfield  {author} {\bibinfo {author} {\bibfnamefont {F.}~\bibnamefont
  {Acernese}} \emph {et~al.} (\bibinfo {collaboration} {VIRGO}),\ }\bibfield
  {title} {\enquote {\bibinfo {title} {{Advanced Virgo: a second-generation
  interferometric gravitational wave detector}},}\ }\href {\doibase
  10.1088/0264-9381/32/2/024001} {\bibfield  {journal} {\bibinfo  {journal}
  {Class. Quant. Grav.}\ }\textbf {\bibinfo {volume} {32}},\ \bibinfo {pages}
  {024001} (\bibinfo {year} {2015})},\ \Eprint {http://arxiv.org/abs/1408.3978}
  {arXiv:1408.3978 [gr-qc]} \BibitemShut {NoStop}%
%%CITATION = ARXIV:1408.3978;%%
\bibitem [{\citenamefont {Aso}\ \emph {et~al.}(2013)\citenamefont {Aso},
  \citenamefont {Michimura}, \citenamefont {Somiya}, \citenamefont {Ando},
  \citenamefont {Miyakawa}, \citenamefont {Sekiguchi}, \citenamefont
  {Tatsumi},\ and\ \citenamefont {Yamamoto}}]{KAGRADet}%
  \BibitemOpen
  \bibfield  {author} {\bibinfo {author} {\bibfnamefont {Yoichi}\ \bibnamefont
  {Aso}}, \bibinfo {author} {\bibfnamefont {Yuta}\ \bibnamefont {Michimura}},
  \bibinfo {author} {\bibfnamefont {Kentaro}\ \bibnamefont {Somiya}}, \bibinfo
  {author} {\bibfnamefont {Masaki}\ \bibnamefont {Ando}}, \bibinfo {author}
  {\bibfnamefont {Osamu}\ \bibnamefont {Miyakawa}}, \bibinfo {author}
  {\bibfnamefont {Takanori}\ \bibnamefont {Sekiguchi}}, \bibinfo {author}
  {\bibfnamefont {Daisuke}\ \bibnamefont {Tatsumi}}, \ and\ \bibinfo {author}
  {\bibfnamefont {Hiroaki}\ \bibnamefont {Yamamoto}} (\bibinfo {collaboration}
  {KAGRA}),\ }\bibfield  {title} {\enquote {\bibinfo {title} {{Interferometer
  design of the KAGRA gravitational wave detector}},}\ }\href {\doibase
  10.1103/PhysRevD.88.043007} {\bibfield  {journal} {\bibinfo  {journal} {Phys.
  Rev.}\ }\textbf {\bibinfo {volume} {D88}},\ \bibinfo {pages} {043007}
  (\bibinfo {year} {2013})},\ \Eprint {http://arxiv.org/abs/1306.6747}
  {arXiv:1306.6747 [gr-qc]} \BibitemShut {NoStop}%
%%CITATION = ARXIV:1306.6747;%%
\bibitem [{\citenamefont {Abbott}\ \emph
  {et~al.}(2018{\natexlab{a}})\citenamefont {Abbott} \emph
  {et~al.}}]{Aasi:2013wya}%
  \BibitemOpen
  \bibfield  {author} {\bibinfo {author} {\bibfnamefont {B.~P.}\ \bibnamefont
  {Abbott}} \emph {et~al.} (\bibinfo {collaboration} {KAGRA, LIGO Scientific,
  VIRGO}),\ }\bibfield  {title} {\enquote {\bibinfo {title} {{Prospects for
  Observing and Localizing Gravitational-Wave Transients with Advanced LIGO,
  Advanced Virgo and KAGRA}},}\ }\href {\doibase 10.1007/s41114-018-0012-9,
  10.1007/lrr-2016-1} {\bibfield  {journal} {\bibinfo  {journal} {Living Rev.
  Rel.}\ }\textbf {\bibinfo {volume} {21}},\ \bibinfo {pages} {3} (\bibinfo
  {year} {2018}{\natexlab{a}})},\ \Eprint {http://arxiv.org/abs/1304.0670}
  {arXiv:1304.0670 [gr-qc]} \BibitemShut {NoStop}%
%%CITATION = ARXIV:1304.0670;%%
\bibitem [{\citenamefont {Iyer}\ \emph {et~al.}(2011)\citenamefont {Iyer} \emph
  {et~al.}}]{LI-Det}%
  \BibitemOpen
  \bibfield  {author} {\bibinfo {author} {\bibfnamefont {Bala.}\ \bibnamefont
  {Iyer}} \emph {et~al.},\ }\href@noop {} {\enquote {\bibinfo {title}
  {{LIGO-India Technical Report No. LIGOM1100296}},}\ } (\bibinfo {year}
  {2011})\BibitemShut {NoStop}%
\bibitem [{\citenamefont {Saleem}\ \emph {et~al.}(2022)\citenamefont {Saleem}
  \emph {et~al.}}]{Saleem:2021iwi}%
  \BibitemOpen
  \bibfield  {author} {\bibinfo {author} {\bibfnamefont {M.}~\bibnamefont
  {Saleem}} \emph {et~al.},\ }\bibfield  {title} {\enquote {\bibinfo {title}
  {{The science case for LIGO-India}},}\ }\href {\doibase
  10.1088/1361-6382/ac3b99} {\bibfield  {journal} {\bibinfo  {journal} {Class.
  Quant. Grav.}\ }\textbf {\bibinfo {volume} {39}},\ \bibinfo {pages} {025004}
  (\bibinfo {year} {2022})},\ \Eprint {http://arxiv.org/abs/2105.01716}
  {arXiv:2105.01716 [gr-qc]} \BibitemShut {NoStop}%
\bibitem [{\citenamefont {Abbott}\ \emph
  {et~al.}(2018{\natexlab{b}})\citenamefont {Abbott} \emph {et~al.}}]{A+}%
  \BibitemOpen
  \bibfield  {author} {\bibinfo {author} {\bibfnamefont {B.~P.}\ \bibnamefont
  {Abbott}} \emph {et~al.} (\bibinfo {collaboration} {KAGRA, LIGO Scientific,
  VIRGO}),\ }\bibfield  {title} {\enquote {\bibinfo {title} {{Prospects for
  Observing and Localizing Gravitational-Wave Transients with Advanced LIGO,
  Advanced Virgo and KAGRA}},}\ }\href {\doibase 10.1007/s41114-018-0012-9,
  10.1007/lrr-2016-1} {\bibfield  {journal} {\bibinfo  {journal} {Living Rev.
  Rel.}\ }\textbf {\bibinfo {volume} {21}},\ \bibinfo {pages} {3} (\bibinfo
  {year} {2018}{\natexlab{b}})},\ \Eprint {http://arxiv.org/abs/1304.0670}
  {arXiv:1304.0670 [gr-qc]} \BibitemShut {NoStop}%
%%CITATION = ARXIV:1304.0670;%%
\bibitem [{\citenamefont {{LIGO Scientific
  Collaboration}}(2019)}]{LIGO:ISWP2019}%
  \BibitemOpen
  \bibfield  {author} {\bibinfo {author} {\bibnamefont {{LIGO Scientific
  Collaboration}}},\ }\href@noop {} {\enquote {\bibinfo {title} {{Instrument
  Science White Paper 2019}},}\ } (\bibinfo {year} {2019}),\ \Eprint
  {http://arxiv.org/abs/{LIGO-T1900409}} {{LIGO-T1900409}} \BibitemShut
  {NoStop}%
\bibitem [{\citenamefont {Adhikari}\ \emph {et~al.}(2019)\citenamefont
  {Adhikari}, \citenamefont {Ajith}, \citenamefont {Chen}, \citenamefont
  {Clark}, \citenamefont {Dergachev}, \citenamefont {Fotopoulos}, \citenamefont
  {Gossan}, \citenamefont {Mandel}, \citenamefont {Okounkova}, \citenamefont
  {Raymond},\ and\ \citenamefont {Read}}]{Adhikari_2019}%
  \BibitemOpen
  \bibfield  {author} {\bibinfo {author} {\bibfnamefont {R~X}\ \bibnamefont
  {Adhikari}}, \bibinfo {author} {\bibfnamefont {P}~\bibnamefont {Ajith}},
  \bibinfo {author} {\bibfnamefont {Y}~\bibnamefont {Chen}}, \bibinfo {author}
  {\bibfnamefont {J~A}\ \bibnamefont {Clark}}, \bibinfo {author} {\bibfnamefont
  {V}~\bibnamefont {Dergachev}}, \bibinfo {author} {\bibfnamefont {N~V}\
  \bibnamefont {Fotopoulos}}, \bibinfo {author} {\bibfnamefont {S~E}\
  \bibnamefont {Gossan}}, \bibinfo {author} {\bibfnamefont {I}~\bibnamefont
  {Mandel}}, \bibinfo {author} {\bibfnamefont {M}~\bibnamefont {Okounkova}},
  \bibinfo {author} {\bibfnamefont {V}~\bibnamefont {Raymond}}, \ and\ \bibinfo
  {author} {\bibfnamefont {J~S}\ \bibnamefont {Read}},\ }\bibfield  {title}
  {\enquote {\bibinfo {title} {Astrophysical science metrics for
  next-generation gravitational-wave detectors},}\ }\href {\doibase
  10.1088/1361-6382/ab3cff} {\bibfield  {journal} {\bibinfo  {journal}
  {Classical and Quantum Gravity}\ }\textbf {\bibinfo {volume} {36}},\ \bibinfo
  {pages} {245010} (\bibinfo {year} {2019})}\BibitemShut {NoStop}%
\bibitem [{\citenamefont {Punturo}\ \emph {et~al.}(2010)\citenamefont {Punturo}
  \emph {et~al.}}]{Punturo:2010zz}%
  \BibitemOpen
  \bibfield  {author} {\bibinfo {author} {\bibfnamefont {M.}~\bibnamefont
  {Punturo}} \emph {et~al.},\ }\bibfield  {title} {\enquote {\bibinfo {title}
  {{The Einstein Telescope: A third-generation gravitational wave
  observatory}},}\ }\bibfield  {booktitle} {\emph {\bibinfo {booktitle}
  {{Proceedings, 14th Workshop on Gravitational wave data analysis (GWDAW-14):
  Rome, Italy, January 26-29, 2010}}},\ }\href {\doibase
  10.1088/0264-9381/27/19/194002} {\bibfield  {journal} {\bibinfo  {journal}
  {Class. Quant. Grav.}\ }\textbf {\bibinfo {volume} {27}},\ \bibinfo {pages}
  {194002} (\bibinfo {year} {2010})}\BibitemShut {NoStop}%
%%CITATION = CQGRD,27,194002;%%
\bibitem [{\citenamefont {Dwyer}\ \emph {et~al.}(2015)\citenamefont {Dwyer},
  \citenamefont {Sigg}, \citenamefont {Ballmer}, \citenamefont {Barsotti},
  \citenamefont {Mavalvala},\ and\ \citenamefont {Evans}}]{Dwyer_2015}%
  \BibitemOpen
  \bibfield  {author} {\bibinfo {author} {\bibfnamefont {Sheila}\ \bibnamefont
  {Dwyer}}, \bibinfo {author} {\bibfnamefont {Daniel}\ \bibnamefont {Sigg}},
  \bibinfo {author} {\bibfnamefont {Stefan~W.}\ \bibnamefont {Ballmer}},
  \bibinfo {author} {\bibfnamefont {Lisa}\ \bibnamefont {Barsotti}}, \bibinfo
  {author} {\bibfnamefont {Nergis}\ \bibnamefont {Mavalvala}}, \ and\ \bibinfo
  {author} {\bibfnamefont {Matthew}\ \bibnamefont {Evans}},\ }\bibfield
  {title} {\enquote {\bibinfo {title} {Gravitational wave detector with
  cosmological reach},}\ }\href {\doibase 10.1103/physrevd.91.082001}
  {\bibfield  {journal} {\bibinfo  {journal} {Physical Review D}\ }\textbf
  {\bibinfo {volume} {91}} (\bibinfo {year} {2015}),\
  10.1103/physrevd.91.082001}\BibitemShut {NoStop}%
\bibitem [{\citenamefont {Vitale}\ and\ \citenamefont
  {Evans}(2017)}]{Vitale:2016icu}%
  \BibitemOpen
  \bibfield  {author} {\bibinfo {author} {\bibfnamefont {Salvatore}\
  \bibnamefont {Vitale}}\ and\ \bibinfo {author} {\bibfnamefont {Matthew}\
  \bibnamefont {Evans}},\ }\bibfield  {title} {\enquote {\bibinfo {title}
  {{Parameter estimation for binary black holes with networks of third
  generation gravitational-wave detectors}},}\ }\href {\doibase
  10.1103/PhysRevD.95.064052} {\bibfield  {journal} {\bibinfo  {journal} {Phys.
  Rev.}\ }\textbf {\bibinfo {volume} {D95}},\ \bibinfo {pages} {064052}
  (\bibinfo {year} {2017})},\ \Eprint {http://arxiv.org/abs/1610.06917}
  {arXiv:1610.06917 [gr-qc]} \BibitemShut {NoStop}%
%%CITATION = ARXIV:1610.06917;%%
\bibitem [{\citenamefont {Abbott}\ \emph
  {et~al.}(2017{\natexlab{a}})\citenamefont {Abbott} \emph
  {et~al.}}]{GBM:2017lvd}%
  \BibitemOpen
  \bibfield  {author} {\bibinfo {author} {\bibfnamefont {B.~P.}\ \bibnamefont
  {Abbott}} \emph {et~al.},\ }\bibfield  {title} {\enquote {\bibinfo {title}
  {{Multi-messenger Observations of a Binary Neutron Star Merger}},}\ }\href
  {\doibase 10.3847/2041-8213/aa91c9} {\bibfield  {journal} {\bibinfo
  {journal} {Astrophys. J.}\ }\textbf {\bibinfo {volume} {848}},\ \bibinfo
  {pages} {L12} (\bibinfo {year} {2017}{\natexlab{a}})},\ \Eprint
  {http://arxiv.org/abs/1710.05833} {arXiv:1710.05833 [astro-ph.HE]}
  \BibitemShut {NoStop}%
%%CITATION = ARXIV:1710.05833;%%
\bibitem [{\citenamefont {Goldstein}\ \emph {et~al.}(2017)\citenamefont
  {Goldstein} \emph {et~al.}}]{Goldstein:2017mmi}%
  \BibitemOpen
  \bibfield  {author} {\bibinfo {author} {\bibfnamefont {A.}~\bibnamefont
  {Goldstein}} \emph {et~al.},\ }\bibfield  {title} {\enquote {\bibinfo {title}
  {{An Ordinary Short Gamma-Ray Burst with Extraordinary Implications:
  Fermi-GBM Detection of GRB 170817A}},}\ }\href {\doibase
  10.3847/2041-8213/aa8f41} {\bibfield  {journal} {\bibinfo  {journal}
  {Astrophys. J. Lett.}\ }\textbf {\bibinfo {volume} {848}},\ \bibinfo {pages}
  {L14} (\bibinfo {year} {2017})},\ \Eprint {http://arxiv.org/abs/1710.05446}
  {arXiv:1710.05446 [astro-ph.HE]} \BibitemShut {NoStop}%
\bibitem [{\citenamefont {Abbott}\ \emph
  {et~al.}(2017{\natexlab{b}})\citenamefont {Abbott} \emph
  {et~al.}}]{LIGOScientific:2017zic}%
  \BibitemOpen
  \bibfield  {author} {\bibinfo {author} {\bibfnamefont {B.~P.}\ \bibnamefont
  {Abbott}} \emph {et~al.} (\bibinfo {collaboration} {LIGO Scientific, Virgo,
  Fermi-GBM, INTEGRAL}),\ }\bibfield  {title} {\enquote {\bibinfo {title}
  {{Gravitational Waves and Gamma-rays from a Binary Neutron Star Merger:
  GW170817 and GRB 170817A}},}\ }\href {\doibase 10.3847/2041-8213/aa920c}
  {\bibfield  {journal} {\bibinfo  {journal} {Astrophys. J. Lett.}\ }\textbf
  {\bibinfo {volume} {848}},\ \bibinfo {pages} {L13} (\bibinfo {year}
  {2017}{\natexlab{b}})},\ \Eprint {http://arxiv.org/abs/1710.05834}
  {arXiv:1710.05834 [astro-ph.HE]} \BibitemShut {NoStop}%
\bibitem [{\citenamefont {Savchenko}\ \emph {et~al.}(2017)\citenamefont
  {Savchenko} \emph {et~al.}}]{Savchenko:2017ffs}%
  \BibitemOpen
  \bibfield  {author} {\bibinfo {author} {\bibfnamefont {V.}~\bibnamefont
  {Savchenko}} \emph {et~al.},\ }\bibfield  {title} {\enquote {\bibinfo {title}
  {{INTEGRAL Detection of the First Prompt Gamma-Ray Signal Coincident with the
  Gravitational-wave Event GW170817}},}\ }\href {\doibase
  10.3847/2041-8213/aa8f94} {\bibfield  {journal} {\bibinfo  {journal}
  {Astrophys. J. Lett.}\ }\textbf {\bibinfo {volume} {848}},\ \bibinfo {pages}
  {L15} (\bibinfo {year} {2017})},\ \Eprint {http://arxiv.org/abs/1710.05449}
  {arXiv:1710.05449 [astro-ph.HE]} \BibitemShut {NoStop}%
\bibitem [{\citenamefont {Abbott}\ \emph
  {et~al.}(2017{\natexlab{c}})\citenamefont {Abbott} \emph
  {et~al.}}]{Abbott:2017xzu}%
  \BibitemOpen
  \bibfield  {author} {\bibinfo {author} {\bibfnamefont {B.~P.}\ \bibnamefont
  {Abbott}} \emph {et~al.} (\bibinfo {collaboration} {LIGO Scientific, Virgo,
  1M2H, Dark Energy Camera GW-E, DES, DLT40, Las Cumbres Observatory, VINROUGE,
  MASTER}),\ }\bibfield  {title} {\enquote {\bibinfo {title} {{A
  gravitational-wave standard siren measurement of the Hubble constant}},}\
  }\href {\doibase 10.1038/nature24471} {\bibfield  {journal} {\bibinfo
  {journal} {Nature}\ }\textbf {\bibinfo {volume} {551}},\ \bibinfo {pages}
  {85--88} (\bibinfo {year} {2017}{\natexlab{c}})},\ \Eprint
  {http://arxiv.org/abs/1710.05835} {arXiv:1710.05835 [astro-ph.CO]}
  \BibitemShut {NoStop}%
%%CITATION = ARXIV:1710.05835;%%
\bibitem [{\citenamefont {Aghanim}\ \emph {et~al.}(2020)\citenamefont {Aghanim}
  \emph {et~al.}}]{Aghanim:2018eyx}%
  \BibitemOpen
  \bibfield  {author} {\bibinfo {author} {\bibfnamefont {N.}~\bibnamefont
  {Aghanim}} \emph {et~al.} (\bibinfo {collaboration} {Planck}),\ }\bibfield
  {title} {\enquote {\bibinfo {title} {{Planck 2018 results. VI. Cosmological
  parameters}},}\ }\href {\doibase 10.1051/0004-6361/201833910} {\bibfield
  {journal} {\bibinfo  {journal} {Astron. Astrophys.}\ }\textbf {\bibinfo
  {volume} {641}},\ \bibinfo {pages} {A6} (\bibinfo {year} {2020})},\ \bibinfo
  {note} {[Erratum: Astron.Astrophys. 652, C4 (2021)]},\ \Eprint
  {http://arxiv.org/abs/1807.06209} {arXiv:1807.06209 [astro-ph.CO]}
  \BibitemShut {NoStop}%
\bibitem [{\citenamefont {Riess}\ \emph {et~al.}(2016)\citenamefont {Riess}
  \emph {et~al.}}]{Riess:2016jrr}%
  \BibitemOpen
  \bibfield  {author} {\bibinfo {author} {\bibfnamefont {Adam~G.}\ \bibnamefont
  {Riess}} \emph {et~al.},\ }\bibfield  {title} {\enquote {\bibinfo {title} {{A
  2.4\% Determination of the Local Value of the Hubble Constant}},}\ }\href
  {\doibase 10.3847/0004-637X/826/1/56} {\bibfield  {journal} {\bibinfo
  {journal} {Astrophys. J.}\ }\textbf {\bibinfo {volume} {826}},\ \bibinfo
  {pages} {56} (\bibinfo {year} {2016})},\ \Eprint
  {http://arxiv.org/abs/1604.01424} {arXiv:1604.01424 [astro-ph.CO]}
  \BibitemShut {NoStop}%
%%CITATION = ARXIV:1604.01424;%%
\bibitem [{\citenamefont {Schutz}(1986)}]{Schutz:1986gp}%
  \BibitemOpen
  \bibfield  {author} {\bibinfo {author} {\bibfnamefont {Bernard~F.}\
  \bibnamefont {Schutz}},\ }\bibfield  {title} {\enquote {\bibinfo {title}
  {{Determining the Hubble Constant from Gravitational Wave Observations}},}\
  }\href {\doibase 10.1038/323310a0} {\bibfield  {journal} {\bibinfo  {journal}
  {Nature}\ }\textbf {\bibinfo {volume} {323}},\ \bibinfo {pages} {310--311}
  (\bibinfo {year} {1986})}\BibitemShut {NoStop}%
%%CITATION = NATUA,323,310;%%
\bibitem [{\citenamefont {Del~Pozzo}(2012)}]{DelPozzo:2011yh}%
  \BibitemOpen
  \bibfield  {author} {\bibinfo {author} {\bibfnamefont {Walter}\ \bibnamefont
  {Del~Pozzo}},\ }\bibfield  {title} {\enquote {\bibinfo {title} {{Inference of
  the cosmological parameters from gravitational waves: application to second
  generation interferometers}},}\ }\href {\doibase 10.1103/PhysRevD.86.043011}
  {\bibfield  {journal} {\bibinfo  {journal} {Phys. Rev.}\ }\textbf {\bibinfo
  {volume} {D86}},\ \bibinfo {pages} {043011} (\bibinfo {year} {2012})},\
  \Eprint {http://arxiv.org/abs/1108.1317} {arXiv:1108.1317 [astro-ph.CO]}
  \BibitemShut {NoStop}%
%%CITATION = ARXIV:1108.1317;%%
\bibitem [{\citenamefont {Chen}\ \emph {et~al.}(2018)\citenamefont {Chen},
  \citenamefont {Fishbach},\ and\ \citenamefont {Holz}}]{Chen:2017rfc}%
  \BibitemOpen
  \bibfield  {author} {\bibinfo {author} {\bibfnamefont {Hsin-Yu}\ \bibnamefont
  {Chen}}, \bibinfo {author} {\bibfnamefont {Maya}\ \bibnamefont {Fishbach}}, \
  and\ \bibinfo {author} {\bibfnamefont {Daniel~E.}\ \bibnamefont {Holz}},\
  }\bibfield  {title} {\enquote {\bibinfo {title} {{A two per cent Hubble
  constant measurement from standard sirens within five years}},}\ }\href
  {\doibase 10.1038/s41586-018-0606-0} {\bibfield  {journal} {\bibinfo
  {journal} {Nature}\ }\textbf {\bibinfo {volume} {562}},\ \bibinfo {pages}
  {545--547} (\bibinfo {year} {2018})},\ \Eprint
  {http://arxiv.org/abs/1712.06531} {arXiv:1712.06531 [astro-ph.CO]}
  \BibitemShut {NoStop}%
\bibitem [{\citenamefont {Fishbach}\ \emph {et~al.}(2019)\citenamefont
  {Fishbach} \emph {et~al.}}]{Fishbach:2018gjp}%
  \BibitemOpen
  \bibfield  {author} {\bibinfo {author} {\bibfnamefont {M.}~\bibnamefont
  {Fishbach}} \emph {et~al.} (\bibinfo {collaboration} {LIGO Scientific,
  Virgo}),\ }\bibfield  {title} {\enquote {\bibinfo {title} {{A Standard Siren
  Measurement of the Hubble Constant from GW170817 without the Electromagnetic
  Counterpart}},}\ }\href {\doibase 10.3847/2041-8213/aaf96e} {\bibfield
  {journal} {\bibinfo  {journal} {Astrophys. J.}\ }\textbf {\bibinfo {volume}
  {871}},\ \bibinfo {pages} {L13} (\bibinfo {year} {2019})},\ \Eprint
  {http://arxiv.org/abs/1807.05667} {arXiv:1807.05667 [astro-ph.CO]}
  \BibitemShut {NoStop}%
%%CITATION = ARXIV:1807.05667;%%
\bibitem [{\citenamefont {Nair}\ \emph {et~al.}(2018)\citenamefont {Nair},
  \citenamefont {Bose},\ and\ \citenamefont {Saini}}]{Nair:2018ign}%
  \BibitemOpen
  \bibfield  {author} {\bibinfo {author} {\bibfnamefont {Remya}\ \bibnamefont
  {Nair}}, \bibinfo {author} {\bibfnamefont {Sukanta}\ \bibnamefont {Bose}}, \
  and\ \bibinfo {author} {\bibfnamefont {Tarun~Deep}\ \bibnamefont {Saini}},\
  }\bibfield  {title} {\enquote {\bibinfo {title} {{Measuring the Hubble
  constant: Gravitational wave observations meet galaxy clustering}},}\ }\href
  {\doibase 10.1103/PhysRevD.98.023502} {\bibfield  {journal} {\bibinfo
  {journal} {Phys. Rev.}\ }\textbf {\bibinfo {volume} {D98}},\ \bibinfo {pages}
  {023502} (\bibinfo {year} {2018})},\ \Eprint
  {http://arxiv.org/abs/1804.06085} {arXiv:1804.06085 [astro-ph.CO]}
  \BibitemShut {NoStop}%
%%CITATION = ARXIV:1804.06085;%%
\bibitem [{\citenamefont {Osato}(2018)}]{Osato:2018mtm}%
  \BibitemOpen
  \bibfield  {author} {\bibinfo {author} {\bibfnamefont {Ken}\ \bibnamefont
  {Osato}},\ }\bibfield  {title} {\enquote {\bibinfo {title} {{Exploring the
  distance-redshift relation with gravitational wave standard sirens and
  tomographic weak lensing}},}\ }\href {\doibase 10.1103/PhysRevD.98.083524}
  {\bibfield  {journal} {\bibinfo  {journal} {Phys. Rev.}\ }\textbf {\bibinfo
  {volume} {D98}},\ \bibinfo {pages} {083524} (\bibinfo {year} {2018})},\
  \Eprint {http://arxiv.org/abs/1807.00016} {arXiv:1807.00016 [astro-ph.CO]}
  \BibitemShut {NoStop}%
%%CITATION = ARXIV:1807.00016;%%
\bibitem [{\citenamefont {Soares-Santos}\ \emph {et~al.}(2019)\citenamefont
  {Soares-Santos} \emph {et~al.}}]{Soares-Santos:2019irc}%
  \BibitemOpen
  \bibfield  {author} {\bibinfo {author} {\bibfnamefont {M.}~\bibnamefont
  {Soares-Santos}} \emph {et~al.} (\bibinfo {collaboration} {DES, LIGO
  Scientific, Virgo}),\ }\bibfield  {title} {\enquote {\bibinfo {title} {{First
  Measurement of the Hubble Constant from a Dark Standard Siren using the Dark
  Energy Survey Galaxies and the LIGO/Virgo Binary–Black-hole Merger
  GW170814}},}\ }\href {\doibase 10.3847/2041-8213/ab14f1} {\bibfield
  {journal} {\bibinfo  {journal} {Astrophys. J.}\ }\textbf {\bibinfo {volume}
  {876}},\ \bibinfo {pages} {L7} (\bibinfo {year} {2019})},\ \Eprint
  {http://arxiv.org/abs/1901.01540} {arXiv:1901.01540 [astro-ph.CO]}
  \BibitemShut {NoStop}%
%%CITATION = ARXIV:1901.01540;%%
\bibitem [{\citenamefont {Gray}\ \emph {et~al.}(2020)\citenamefont {Gray} \emph
  {et~al.}}]{Gray:2019ksv}%
  \BibitemOpen
  \bibfield  {author} {\bibinfo {author} {\bibfnamefont {Rachel}\ \bibnamefont
  {Gray}} \emph {et~al.},\ }\bibfield  {title} {\enquote {\bibinfo {title}
  {{Cosmological inference using gravitational wave standard sirens: A mock
  data analysis}},}\ }\href {\doibase 10.1103/PhysRevD.101.122001} {\bibfield
  {journal} {\bibinfo  {journal} {Phys. Rev. D}\ }\textbf {\bibinfo {volume}
  {101}},\ \bibinfo {pages} {122001} (\bibinfo {year} {2020})},\ \Eprint
  {http://arxiv.org/abs/1908.06050} {arXiv:1908.06050 [gr-qc]} \BibitemShut
  {NoStop}%
\bibitem [{\citenamefont {Abbott}\ \emph
  {et~al.}(2021{\natexlab{c}})\citenamefont {Abbott} \emph
  {et~al.}}]{Abbott:2019yzh}%
  \BibitemOpen
  \bibfield  {author} {\bibinfo {author} {\bibfnamefont {B.~P.}\ \bibnamefont
  {Abbott}} \emph {et~al.} (\bibinfo {collaboration} {LIGO Scientific, Virgo,
  VIRGO}),\ }\bibfield  {title} {\enquote {\bibinfo {title} {{A
  Gravitational-wave Measurement of the Hubble Constant Following the Second
  Observing Run of Advanced LIGO and Virgo}},}\ }\href {\doibase
  10.3847/1538-4357/abdcb7} {\bibfield  {journal} {\bibinfo  {journal}
  {Astrophys. J.}\ }\textbf {\bibinfo {volume} {909}},\ \bibinfo {pages} {218}
  (\bibinfo {year} {2021}{\natexlab{c}})},\ \Eprint
  {http://arxiv.org/abs/1908.06060} {arXiv:1908.06060 [astro-ph.CO]}
  \BibitemShut {NoStop}%
\bibitem [{\citenamefont {Namikawa}\ \emph {et~al.}(2016)\citenamefont
  {Namikawa}, \citenamefont {Nishizawa},\ and\ \citenamefont
  {Taruya}}]{Namikawa:2016edr}%
  \BibitemOpen
  \bibfield  {author} {\bibinfo {author} {\bibfnamefont {Toshiya}\ \bibnamefont
  {Namikawa}}, \bibinfo {author} {\bibfnamefont {Atsushi}\ \bibnamefont
  {Nishizawa}}, \ and\ \bibinfo {author} {\bibfnamefont {Atsushi}\ \bibnamefont
  {Taruya}},\ }\bibfield  {title} {\enquote {\bibinfo {title} {{Detecting
  Black-Hole Binary Clustering via the Second-Generation Gravitational-Wave
  Detectors}},}\ }\href {\doibase 10.1103/PhysRevD.94.024013} {\bibfield
  {journal} {\bibinfo  {journal} {Phys. Rev.}\ }\textbf {\bibinfo {volume}
  {D94}},\ \bibinfo {pages} {024013} (\bibinfo {year} {2016})},\ \Eprint
  {http://arxiv.org/abs/1603.08072} {arXiv:1603.08072 [astro-ph.CO]}
  \BibitemShut {NoStop}%
%%CITATION = ARXIV:1603.08072;%%
\bibitem [{\citenamefont {Calore}\ \emph {et~al.}(2020)\citenamefont {Calore},
  \citenamefont {Cuoco}, \citenamefont {Regimbau}, \citenamefont {Sachdev},\
  and\ \citenamefont {Serpico}}]{Calore:2020bpd}%
  \BibitemOpen
  \bibfield  {author} {\bibinfo {author} {\bibfnamefont {Francesca}\
  \bibnamefont {Calore}}, \bibinfo {author} {\bibfnamefont {Alessandro}\
  \bibnamefont {Cuoco}}, \bibinfo {author} {\bibfnamefont {Tania}\ \bibnamefont
  {Regimbau}}, \bibinfo {author} {\bibfnamefont {Surabhi}\ \bibnamefont
  {Sachdev}}, \ and\ \bibinfo {author} {\bibfnamefont {Pasquale~Dario}\
  \bibnamefont {Serpico}},\ }\bibfield  {title} {\enquote {\bibinfo {title}
  {{Cross-correlating galaxy catalogs and gravitational waves: a tomographic
  approach}},}\ }\href {\doibase 10.1103/PhysRevResearch.2.023314} {\bibfield
  {journal} {\bibinfo  {journal} {Phys. Rev. Res.}\ }\textbf {\bibinfo {volume}
  {2}},\ \bibinfo {pages} {023314} (\bibinfo {year} {2020})},\ \Eprint
  {http://arxiv.org/abs/2002.02466} {arXiv:2002.02466 [astro-ph.CO]}
  \BibitemShut {NoStop}%
\bibitem [{\citenamefont {Mukherjee}\ \emph {et~al.}(2020)\citenamefont
  {Mukherjee}, \citenamefont {Wandelt},\ and\ \citenamefont
  {Silk}}]{Mukherjee:2019wcg}%
  \BibitemOpen
  \bibfield  {author} {\bibinfo {author} {\bibfnamefont {Suvodip}\ \bibnamefont
  {Mukherjee}}, \bibinfo {author} {\bibfnamefont {Benjamin~D.}\ \bibnamefont
  {Wandelt}}, \ and\ \bibinfo {author} {\bibfnamefont {Joseph}\ \bibnamefont
  {Silk}},\ }\bibfield  {title} {\enquote {\bibinfo {title} {{Probing the
  theory of gravity with gravitational lensing of gravitational waves and
  galaxy surveys}},}\ }\href {\doibase 10.1093/mnras/staa827} {\bibfield
  {journal} {\bibinfo  {journal} {Mon. Not. Roy. Astron. Soc.}\ }\textbf
  {\bibinfo {volume} {494}},\ \bibinfo {pages} {1956--1970} (\bibinfo {year}
  {2020})},\ \Eprint {http://arxiv.org/abs/1908.08951} {arXiv:1908.08951
  [astro-ph.CO]} \BibitemShut {NoStop}%
\bibitem [{\citenamefont {Kaiser}(1984)}]{Kaiser:1984sw}%
  \BibitemOpen
  \bibfield  {author} {\bibinfo {author} {\bibfnamefont {Nick}\ \bibnamefont
  {Kaiser}},\ }\bibfield  {title} {\enquote {\bibinfo {title} {{On the Spatial
  correlations of Abell clusters}},}\ }\href {\doibase 10.1086/184341}
  {\bibfield  {journal} {\bibinfo  {journal} {Astrophys. J. Lett.}\ }\textbf
  {\bibinfo {volume} {284}},\ \bibinfo {pages} {L9--L12} (\bibinfo {year}
  {1984})}\BibitemShut {NoStop}%
\bibitem [{\citenamefont {Zehavi}\ \emph {et~al.}(2005)\citenamefont {Zehavi}
  \emph {et~al.}}]{Zehavi:2004ii}%
  \BibitemOpen
  \bibfield  {author} {\bibinfo {author} {\bibfnamefont {Idit}\ \bibnamefont
  {Zehavi}} \emph {et~al.} (\bibinfo {collaboration} {SDSS}),\ }\bibfield
  {title} {\enquote {\bibinfo {title} {{The Luminosity and color dependence of
  the galaxy correlation function}},}\ }\href {\doibase 10.1086/431891}
  {\bibfield  {journal} {\bibinfo  {journal} {Astrophys. J.}\ }\textbf
  {\bibinfo {volume} {630}},\ \bibinfo {pages} {1--27} (\bibinfo {year}
  {2005})},\ \Eprint {http://arxiv.org/abs/astro-ph/0408569}
  {arXiv:astro-ph/0408569 [astro-ph]} \BibitemShut {NoStop}%
%%CITATION = ASTRO-PH/0408569;%%
\bibitem [{\citenamefont {Landy}\ and\ \citenamefont
  {Szalay}(1993)}]{Landy:1993yu}%
  \BibitemOpen
  \bibfield  {author} {\bibinfo {author} {\bibfnamefont {Stephen~D.}\
  \bibnamefont {Landy}}\ and\ \bibinfo {author} {\bibfnamefont {Alexander~S.}\
  \bibnamefont {Szalay}},\ }\bibfield  {title} {\enquote {\bibinfo {title}
  {{Bias and variance of angular correlation functions}},}\ }\href {\doibase
  10.1086/172900} {\bibfield  {journal} {\bibinfo  {journal} {Astrophys. J.}\
  }\textbf {\bibinfo {volume} {412}},\ \bibinfo {pages} {64} (\bibinfo {year}
  {1993})}\BibitemShut {NoStop}%
%%CITATION = ASJOA,412,64;%%
\bibitem [{\citenamefont {Eisenstein}\ and\ \citenamefont
  {Hu}(1997)}]{Eisenstein:1997jh}%
  \BibitemOpen
  \bibfield  {author} {\bibinfo {author} {\bibfnamefont {Daniel~J.}\
  \bibnamefont {Eisenstein}}\ and\ \bibinfo {author} {\bibfnamefont {Wayne}\
  \bibnamefont {Hu}},\ }\bibfield  {title} {\enquote {\bibinfo {title} {{Power
  spectra for cold dark matter and its variants}},}\ }\href {\doibase
  10.1086/306640} {\bibfield  {journal} {\bibinfo  {journal} {Astrophys. J.}\
  }\textbf {\bibinfo {volume} {511}},\ \bibinfo {pages} {5} (\bibinfo {year}
  {1997})},\ \Eprint {http://arxiv.org/abs/astro-ph/9710252}
  {arXiv:astro-ph/9710252 [astro-ph]} \BibitemShut {NoStop}%
%%CITATION = ASTRO-PH/9710252;%%
\bibitem [{\citenamefont {Belczynski}\ \emph {et~al.}(2016)\citenamefont
  {Belczynski} \emph {et~al.}}]{Belczynski}%
  \BibitemOpen
  \bibfield  {author} {\bibinfo {author} {\bibfnamefont {K.}~\bibnamefont
  {Belczynski}} \emph {et~al.},\ }\bibfield  {title} {\enquote {\bibinfo
  {title} {{The Effect of Pair-Instability Mass Loss on Black Hole Mergers}},}\
  }\href {\doibase 10.1051/0004-6361/201628980} {\bibfield  {journal} {\bibinfo
   {journal} {Astron. Astrophys.}\ }\textbf {\bibinfo {volume} {594}},\
  \bibinfo {pages} {A97} (\bibinfo {year} {2016})},\ \Eprint
  {http://arxiv.org/abs/1607.03116} {arXiv:1607.03116 [astro-ph.HE]}
  \BibitemShut {NoStop}%
\bibitem [{\citenamefont {Abbott}\ \emph
  {et~al.}(2019{\natexlab{b}})\citenamefont {Abbott} \emph
  {et~al.}}]{LIGO-Rates}%
  \BibitemOpen
  \bibfield  {author} {\bibinfo {author} {\bibfnamefont {B.P.}\ \bibnamefont
  {Abbott}} \emph {et~al.} (\bibinfo {collaboration} {LIGO Scientific,
  Virgo}),\ }\bibfield  {title} {\enquote {\bibinfo {title} {{Binary Black Hole
  Population Properties Inferred from the First and Second Observing Runs of
  Advanced LIGO and Advanced Virgo}},}\ }\href {\doibase
  10.3847/2041-8213/ab3800} {\bibfield  {journal} {\bibinfo  {journal}
  {Astrophys. J.}\ }\textbf {\bibinfo {volume} {882}},\ \bibinfo {pages} {L24}
  (\bibinfo {year} {2019}{\natexlab{b}})},\ \Eprint
  {http://arxiv.org/abs/1811.12940} {arXiv:1811.12940 [astro-ph.HE]}
  \BibitemShut {NoStop}%
\bibitem [{\citenamefont {Agrawal}\ \emph {et~al.}(2017)\citenamefont
  {Agrawal}, \citenamefont {Makiya}, \citenamefont {Chiang}, \citenamefont
  {Jeong}, \citenamefont {Saito},\ and\ \citenamefont
  {Komatsu}}]{Agrawal:2017khv}%
  \BibitemOpen
  \bibfield  {author} {\bibinfo {author} {\bibfnamefont {Aniket}\ \bibnamefont
  {Agrawal}}, \bibinfo {author} {\bibfnamefont {Ryu}\ \bibnamefont {Makiya}},
  \bibinfo {author} {\bibfnamefont {Chi-Ting}\ \bibnamefont {Chiang}}, \bibinfo
  {author} {\bibfnamefont {Donghui}\ \bibnamefont {Jeong}}, \bibinfo {author}
  {\bibfnamefont {Shun}\ \bibnamefont {Saito}}, \ and\ \bibinfo {author}
  {\bibfnamefont {Eiichiro}\ \bibnamefont {Komatsu}},\ }\bibfield  {title}
  {\enquote {\bibinfo {title} {{Generating Log-normal Mock Catalog of Galaxies
  in Redshift Space}},}\ }\href {\doibase 10.1088/1475-7516/2017/10/003}
  {\bibfield  {journal} {\bibinfo  {journal} {JCAP}\ }\textbf {\bibinfo
  {volume} {1710}},\ \bibinfo {pages} {003} (\bibinfo {year} {2017})},\ \Eprint
  {http://arxiv.org/abs/1706.09195} {arXiv:1706.09195 [astro-ph.CO]}
  \BibitemShut {NoStop}%
%%CITATION = ARXIV:1706.09195;%%
\bibitem [{\citenamefont {{Higson}}\ \emph {et~al.}(2019)\citenamefont
  {{Higson}}, \citenamefont {{Handley}}, \citenamefont {{Hobson}},\ and\
  \citenamefont {{Lasenby}}}]{2019S&C....29..891H}%
  \BibitemOpen
  \bibfield  {author} {\bibinfo {author} {\bibfnamefont {Edward}\ \bibnamefont
  {{Higson}}}, \bibinfo {author} {\bibfnamefont {Will}\ \bibnamefont
  {{Handley}}}, \bibinfo {author} {\bibfnamefont {Mike}\ \bibnamefont
  {{Hobson}}}, \ and\ \bibinfo {author} {\bibfnamefont {Anthony}\ \bibnamefont
  {{Lasenby}}},\ }\bibfield  {title} {\enquote {\bibinfo {title} {{Dynamic
  nested sampling: an improved algorithm for parameter estimation and evidence
  calculation}},}\ }\href {\doibase 10.1007/s11222-018-9844-0} {\bibfield
  {journal} {\bibinfo  {journal} {Statistics and Computing}\ }\textbf {\bibinfo
  {volume} {29}},\ \bibinfo {pages} {891--913} (\bibinfo {year} {2019})},\
  \Eprint {http://arxiv.org/abs/1704.03459} {arXiv:1704.03459 [stat.CO]}
  \BibitemShut {NoStop}%
\bibitem [{\citenamefont {Abbott}\ \emph
  {et~al.}(2019{\natexlab{c}})\citenamefont {Abbott} \emph
  {et~al.}}]{LIGOScientific:2018jsj}%
  \BibitemOpen
  \bibfield  {author} {\bibinfo {author} {\bibfnamefont {B.~P.}\ \bibnamefont
  {Abbott}} \emph {et~al.} (\bibinfo {collaboration} {LIGO Scientific,
  Virgo}),\ }\bibfield  {title} {\enquote {\bibinfo {title} {{Binary Black Hole
  Population Properties Inferred from the First and Second Observing Runs of
  Advanced LIGO and Advanced Virgo}},}\ }\href {\doibase
  10.3847/2041-8213/ab3800} {\bibfield  {journal} {\bibinfo  {journal}
  {Astrophys. J.}\ }\textbf {\bibinfo {volume} {882}},\ \bibinfo {pages} {L24}
  (\bibinfo {year} {2019}{\natexlab{c}})},\ \Eprint
  {http://arxiv.org/abs/1811.12940} {arXiv:1811.12940 [astro-ph.HE]}
  \BibitemShut {NoStop}%
%%CITATION = ARXIV:1811.12940;%%
\bibitem [{\citenamefont {Hannam}\ \emph {et~al.}(2014)\citenamefont {Hannam},
  \citenamefont {Schmidt}, \citenamefont {Bohé}, \citenamefont {Haegel},
  \citenamefont {Husa}, \citenamefont {Ohme}, \citenamefont {Pratten},\ and\
  \citenamefont {Pürrer}}]{Hannam:2013oca}%
  \BibitemOpen
  \bibfield  {author} {\bibinfo {author} {\bibfnamefont {Mark}\ \bibnamefont
  {Hannam}}, \bibinfo {author} {\bibfnamefont {Patricia}\ \bibnamefont
  {Schmidt}}, \bibinfo {author} {\bibfnamefont {Alejandro}\ \bibnamefont
  {Bohé}}, \bibinfo {author} {\bibfnamefont {Leïla}\ \bibnamefont {Haegel}},
  \bibinfo {author} {\bibfnamefont {Sascha}\ \bibnamefont {Husa}}, \bibinfo
  {author} {\bibfnamefont {Frank}\ \bibnamefont {Ohme}}, \bibinfo {author}
  {\bibfnamefont {Geraint}\ \bibnamefont {Pratten}}, \ and\ \bibinfo {author}
  {\bibfnamefont {Michael}\ \bibnamefont {Pürrer}},\ }\bibfield  {title}
  {\enquote {\bibinfo {title} {{Simple Model of Complete Precessing
  Black-Hole-Binary Gravitational Waveforms}},}\ }\href {\doibase
  10.1103/PhysRevLett.113.151101} {\bibfield  {journal} {\bibinfo  {journal}
  {Phys. Rev. Lett.}\ }\textbf {\bibinfo {volume} {113}},\ \bibinfo {pages}
  {151101} (\bibinfo {year} {2014})},\ \Eprint {http://arxiv.org/abs/1308.3271}
  {arXiv:1308.3271 [gr-qc]} \BibitemShut {NoStop}%
%%CITATION = ARXIV:1308.3271;%%
\bibitem [{\citenamefont {{LIGO Scientific Collaboration}}(2018)}]{lalsuite}%
  \BibitemOpen
  \bibfield  {author} {\bibinfo {author} {\bibnamefont {{LIGO Scientific
  Collaboration}}},\ }\href {\doibase 10.7935/GT1W-FZ16} {\enquote {\bibinfo
  {title} {{LIGO} {A}lgorithm {L}ibrary - {LALS}uite},}\ }\bibinfo
  {howpublished} {free software (GPL)} (\bibinfo {year} {2018})\BibitemShut
  {NoStop}%
\bibitem [{\citenamefont {Hild}\ \emph {et~al.}(2011)\citenamefont {Hild} \emph
  {et~al.}}]{Hild:2010id}%
  \BibitemOpen
  \bibfield  {author} {\bibinfo {author} {\bibfnamefont {S.}~\bibnamefont
  {Hild}} \emph {et~al.},\ }\bibfield  {title} {\enquote {\bibinfo {title}
  {{Sensitivity Studies for Third-Generation Gravitational Wave
  Observatories}},}\ }\href {\doibase 10.1088/0264-9381/28/9/094013} {\bibfield
   {journal} {\bibinfo  {journal} {Class. Quant. Grav.}\ }\textbf {\bibinfo
  {volume} {28}},\ \bibinfo {pages} {094013} (\bibinfo {year} {2011})},\
  \Eprint {http://arxiv.org/abs/1012.0908} {arXiv:1012.0908 [gr-qc]}
  \BibitemShut {NoStop}%
%%CITATION = ARXIV:1012.0908;%%
\bibitem [{\citenamefont {Abbott}\ \emph
  {et~al.}(2017{\natexlab{d}})\citenamefont {Abbott} \emph
  {et~al.}}]{Evans:2016mbw}%
  \BibitemOpen
  \bibfield  {author} {\bibinfo {author} {\bibfnamefont {Benjamin~P}\
  \bibnamefont {Abbott}} \emph {et~al.} (\bibinfo {collaboration} {LIGO
  Scientific}),\ }\bibfield  {title} {\enquote {\bibinfo {title} {{Exploring
  the Sensitivity of Next Generation Gravitational Wave Detectors}},}\ }\href
  {\doibase 10.1088/1361-6382/aa51f4} {\bibfield  {journal} {\bibinfo
  {journal} {Class. Quant. Grav.}\ }\textbf {\bibinfo {volume} {34}},\ \bibinfo
  {pages} {044001} (\bibinfo {year} {2017}{\natexlab{d}})},\ \Eprint
  {http://arxiv.org/abs/1607.08697} {arXiv:1607.08697 [astro-ph.IM]}
  \BibitemShut {NoStop}%
%%CITATION = ARXIV:1607.08697;%%
\bibitem [{\citenamefont {Biwer}\ \emph {et~al.}(2019)\citenamefont {Biwer},
  \citenamefont {Capano}, \citenamefont {De}, \citenamefont {Cabero},
  \citenamefont {Brown}, \citenamefont {Nitz},\ and\ \citenamefont
  {Raymond}}]{Biwer:2018osg}%
  \BibitemOpen
  \bibfield  {author} {\bibinfo {author} {\bibfnamefont {C.~M.}\ \bibnamefont
  {Biwer}}, \bibinfo {author} {\bibfnamefont {Collin~D.}\ \bibnamefont
  {Capano}}, \bibinfo {author} {\bibfnamefont {Soumi}\ \bibnamefont {De}},
  \bibinfo {author} {\bibfnamefont {Miriam}\ \bibnamefont {Cabero}}, \bibinfo
  {author} {\bibfnamefont {Duncan~A.}\ \bibnamefont {Brown}}, \bibinfo {author}
  {\bibfnamefont {Alexander~H.}\ \bibnamefont {Nitz}}, \ and\ \bibinfo {author}
  {\bibfnamefont {V.}~\bibnamefont {Raymond}},\ }\bibfield  {title} {\enquote
  {\bibinfo {title} {{PyCBC Inference: A Python-based parameter estimation
  toolkit for compact binary coalescence signals}},}\ }\href {\doibase
  10.1088/1538-3873/aaef0b} {\bibfield  {journal} {\bibinfo  {journal} {Publ.
  Astron. Soc. Pac.}\ }\textbf {\bibinfo {volume} {131}},\ \bibinfo {pages}
  {024503} (\bibinfo {year} {2019})},\ \Eprint
  {http://arxiv.org/abs/1807.10312} {arXiv:1807.10312 [astro-ph.IM]}
  \BibitemShut {NoStop}%
%%CITATION = ARXIV:1807.10312;%%
\bibitem [{\citenamefont {{Hall}}\ and\ \citenamefont
  {{Evans}}(2019)}]{2019CQGra..36v5002H}%
  \BibitemOpen
  \bibfield  {author} {\bibinfo {author} {\bibfnamefont {Evan~D.}\ \bibnamefont
  {{Hall}}}\ and\ \bibinfo {author} {\bibfnamefont {Matthew}\ \bibnamefont
  {{Evans}}},\ }\bibfield  {title} {\enquote {\bibinfo {title} {{Metrics for
  next-generation gravitational-wave detectors}},}\ }\href {\doibase
  10.1088/1361-6382/ab41d6} {\bibfield  {journal} {\bibinfo  {journal}
  {Classical and Quantum Gravity}\ }\textbf {\bibinfo {volume} {36}},\ \bibinfo
  {eid} {225002} (\bibinfo {year} {2019})},\ \Eprint
  {http://arxiv.org/abs/1902.09485} {arXiv:1902.09485 [astro-ph.IM]}
  \BibitemShut {NoStop}%
\bibitem [{\citenamefont {Nitz}\ and\ \citenamefont
  {Dal~Canton}(2021)}]{Nitz2021premerger}%
  \BibitemOpen
  \bibfield  {author} {\bibinfo {author} {\bibfnamefont {Alexander~H.}\
  \bibnamefont {Nitz}}\ and\ \bibinfo {author} {\bibfnamefont {Tito}\
  \bibnamefont {Dal~Canton}},\ }\bibfield  {title} {\enquote {\bibinfo {title}
  {Pre-merger localization of compact-binary mergers with third-generation
  observatories},}\ }\href {\doibase 10.3847/2041-8213/ac1a75} {\bibfield
  {journal} {\bibinfo  {journal} {The Astrophysical Journal Letters}\ }\textbf
  {\bibinfo {volume} {917}},\ \bibinfo {pages} {L27} (\bibinfo {year}
  {2021})}\BibitemShut {NoStop}%
\bibitem [{\citenamefont {Kumar}\ \emph {et~al.}(2022)\citenamefont {Kumar},
  \citenamefont {Vijaykumar},\ and\ \citenamefont {Nitz}}]{Kumar:2021aog}%
  \BibitemOpen
  \bibfield  {author} {\bibinfo {author} {\bibfnamefont {Sumit}\ \bibnamefont
  {Kumar}}, \bibinfo {author} {\bibfnamefont {Aditya}\ \bibnamefont
  {Vijaykumar}}, \ and\ \bibinfo {author} {\bibfnamefont {Alexander~H.}\
  \bibnamefont {Nitz}},\ }\bibfield  {title} {\enquote {\bibinfo {title}
  {{Detecting Baryon Acoustic Oscillations with Third-generation Gravitational
  Wave Observatories}},}\ }\href {\doibase 10.3847/1538-4357/ac5e34} {\bibfield
   {journal} {\bibinfo  {journal} {Astrophys. J.}\ }\textbf {\bibinfo {volume}
  {930}},\ \bibinfo {pages} {113} (\bibinfo {year} {2022})},\ \Eprint
  {http://arxiv.org/abs/2110.06152} {arXiv:2110.06152 [astro-ph.CO]}
  \BibitemShut {NoStop}%
\bibitem [{\citenamefont {Cooray}\ and\ \citenamefont
  {Sheth}(2002)}]{Cooray:2002dia}%
  \BibitemOpen
  \bibfield  {author} {\bibinfo {author} {\bibfnamefont {Asantha}\ \bibnamefont
  {Cooray}}\ and\ \bibinfo {author} {\bibfnamefont {Ravi~K.}\ \bibnamefont
  {Sheth}},\ }\bibfield  {title} {\enquote {\bibinfo {title} {{Halo Models of
  Large Scale Structure}},}\ }\href {\doibase 10.1016/S0370-1573(02)00276-4}
  {\bibfield  {journal} {\bibinfo  {journal} {Phys. Rept.}\ }\textbf {\bibinfo
  {volume} {372}},\ \bibinfo {pages} {1--129} (\bibinfo {year} {2002})},\
  \Eprint {http://arxiv.org/abs/astro-ph/0206508} {arXiv:astro-ph/0206508
  [astro-ph]} \BibitemShut {NoStop}%
%%CITATION = ASTRO-PH/0206508;%%
\bibitem [{\citenamefont {Robitaille}\ \emph {et~al.}(2013)\citenamefont
  {Robitaille} \emph {et~al.}}]{Robitaille:2013mpa}%
  \BibitemOpen
  \bibfield  {author} {\bibinfo {author} {\bibfnamefont {Thomas~P.}\
  \bibnamefont {Robitaille}} \emph {et~al.} (\bibinfo {collaboration}
  {Astropy}),\ }\bibfield  {title} {\enquote {\bibinfo {title} {{Astropy: A
  Community Python Package for Astronomy}},}\ }\href {\doibase
  10.1051/0004-6361/201322068} {\bibfield  {journal} {\bibinfo  {journal}
  {Astron. Astrophys.}\ }\textbf {\bibinfo {volume} {558}},\ \bibinfo {pages}
  {A33} (\bibinfo {year} {2013})},\ \Eprint {http://arxiv.org/abs/1307.6212}
  {arXiv:1307.6212 [astro-ph.IM]} \BibitemShut {NoStop}%
\bibitem [{\citenamefont {Price-Whelan}\ \emph {et~al.}(2018)\citenamefont
  {Price-Whelan} \emph {et~al.}}]{Price-Whelan:2018hus}%
  \BibitemOpen
  \bibfield  {author} {\bibinfo {author} {\bibfnamefont {A.M.}\ \bibnamefont
  {Price-Whelan}} \emph {et~al.},\ }\bibfield  {title} {\enquote {\bibinfo
  {title} {{The Astropy Project: Building an Open-science Project and Status of
  the v2.0 Core Package}},}\ }\href {\doibase 10.3847/1538-3881/aabc4f}
  {\bibfield  {journal} {\bibinfo  {journal} {Astron. J.}\ }\textbf {\bibinfo
  {volume} {156}},\ \bibinfo {pages} {123} (\bibinfo {year} {2018})},\ \Eprint
  {http://arxiv.org/abs/1801.02634} {arXiv:1801.02634} \BibitemShut {NoStop}%
\bibitem [{\citenamefont {Harris}\ \emph {et~al.}(2020)\citenamefont {Harris}
  \emph {et~al.}}]{Harris:2020xlr}%
  \BibitemOpen
  \bibfield  {author} {\bibinfo {author} {\bibfnamefont {Charles~R.}\
  \bibnamefont {Harris}} \emph {et~al.},\ }\bibfield  {title} {\enquote
  {\bibinfo {title} {{Array programming with NumPy}},}\ }\href {\doibase
  10.1038/s41586-020-2649-2} {\bibfield  {journal} {\bibinfo  {journal}
  {Nature}\ }\textbf {\bibinfo {volume} {585}},\ \bibinfo {pages} {357--362}
  (\bibinfo {year} {2020})},\ \Eprint {http://arxiv.org/abs/2006.10256}
  {arXiv:2006.10256 [cs.MS]} \BibitemShut {NoStop}%
\bibitem [{\citenamefont {Virtanen}\ \emph {et~al.}(2020)\citenamefont
  {Virtanen} \emph {et~al.}}]{Virtanen:2019joe}%
  \BibitemOpen
  \bibfield  {author} {\bibinfo {author} {\bibfnamefont {Pauli}\ \bibnamefont
  {Virtanen}} \emph {et~al.},\ }\bibfield  {title} {\enquote {\bibinfo {title}
  {{SciPy 1.0--Fundamental Algorithms for Scientific Computing in Python}},}\
  }\href {\doibase 10.1038/s41592-019-0686-2} {\bibfield  {journal} {\bibinfo
  {journal} {Nature Meth.}\ } (\bibinfo {year} {2020}),\
  10.1038/s41592-019-0686-2},\ \Eprint {http://arxiv.org/abs/1907.10121}
  {arXiv:1907.10121 [cs.MS]} \BibitemShut {NoStop}%
\bibitem [{\citenamefont {Hunter}(2007)}]{Hunter:2007}%
  \BibitemOpen
  \bibfield  {author} {\bibinfo {author} {\bibfnamefont {J.~D.}\ \bibnamefont
  {Hunter}},\ }\bibfield  {title} {\enquote {\bibinfo {title} {Matplotlib: A 2d
  graphics environment},}\ }\href {\doibase 10.1109/MCSE.2007.55} {\bibfield
  {journal} {\bibinfo  {journal} {Computing in Science \& Engineering}\
  }\textbf {\bibinfo {volume} {9}},\ \bibinfo {pages} {90--95} (\bibinfo {year}
  {2007})}\BibitemShut {NoStop}%
\end{thebibliography}%

\end{document}